\documentclass[spanish,american,aps,twocolumn]{revtex4-1}
\usepackage[T1]{fontenc}
\usepackage[latin9]{inputenc}
\usepackage{verbatim}
\usepackage{amsmath}
\usepackage{amssymb}
\usepackage{wasysym}
\usepackage{graphicx}
\usepackage{esint}

\makeatletter

\providecommand{\tabularnewline}{\\}

\@ifundefined{textcolor}{}
{%
 \definecolor{BLACK}{gray}{0}
 \definecolor{WHITE}{gray}{1}
 \definecolor{RED}{rgb}{1,0,0}
 \definecolor{GREEN}{rgb}{0,1,0}
 \definecolor{BLUE}{rgb}{0,0,1}
 \definecolor{CYAN}{cmyk}{1,0,0,0}
 \definecolor{MAGENTA}{cmyk}{0,1,0,0}
 \definecolor{YELLOW}{cmyk}{0,0,1,0}
}

\makeatother

\usepackage{babel}
\addto\shorthandsspanish{\spanishdeactivate{~<>}}

\begin{document}

\title{Structure, thermodynamic properties, and phase diagrams of few colloids
confined in a spherical pore}

\author{$^{*\dagger}$Iván E. Paganini}

\author{$^{*\dagger}$Claudio Pastorino}

\email{pastor@cnea.gov.ar }

\author{$^{*\dagger}$Ignacio Urrutia}

\email{iurrutia@cnea.gov.ar}

\affiliation{$^{*}$Departamento de Física de la Materia Condensada, Centro Atómico
Constituyentes, CNEA, Av.Gral.~Paz 1499, 1650 Pcia.~de Buenos Aires,
Argentina}

\affiliation{$^{\dagger}$CONICET, Avenida Rivadavia 1917, C1033AAJ Buenos Aires,
Argentina}
\begin{abstract}
We study a system of few colloids confined in a small spherical cavity
by event driven molecular dynamics simulations in the canonical ensemble.
The colloidal particles interact through a short range square-well
potential, which takes into account the basic elements of attraction
and excluded-volume repulsion of the interaction among colloids. We
analyze the structural and thermodynamic properties of this few-body
confined system in the framework of the theory of inhomogeneous fluids.
Pair correlation functions and density profiles across the cavity
are used to determine the structure of the system and the spatial
characteristics of its inhomogeneities. Pressure on the walls, internal
energy and surface quantities such as surface tension and adsorption
are also analyzed for the whole range of densities, temperatures and
number of particles considered. We have characterized the structure
of systems from 2 to 6 confined particles as function of density and
temperature, identifying the distinctive qualitative behaviors all
over the thermodynamic plane $T-\rho$ in a few-particle equivalence
to phase diagrams of macroscopic systems. Applying the extended law
of corresponding states the square well interaction is mapped to the
Asakura-Oosawa model for colloid-polymer mixtures. We link explicitly
the temperature in the confined square-well fluid to the equivalent
packing fraction of polymers in the Asakura-Oosawa model. Using this
approach we study the confined system of few colloids in a colloid-polymer
mixture.
\end{abstract}
\maketitle

\section{Introduction\label{sec:Intro}}

During last decades colloid physics has been one of the areas of more
intensive activity in the field of soft matter. Colloidal dispersions
have wide and well known applications in the chemical, pharmaceutical,
medical and food industries.\cite{Wang_11} Moreover, colloids posses
certain properties that make them good model systems for basic research.\cite{Weeks_2012,Sacanna_2010,Meng_2010,Lu_2008}
The size of colloidal particles, in the range \foreignlanguage{spanish}{$1nm-10\mu m$},
make possible its direct experimental observation giving rise to beautiful
and conclusive experiments. It has enabled the direct study of phase
transitions like the gas-liquid, the gas-solid and gelation ones,
which is impossible for simple fluids.\cite{Wang_2012,Aarts_2004_b,Anderson_2002,deHoog_2001}

Some colloidal suspensions behave like hard spheres (HS) where the
unique relevant feature of the interacting potential is the repulsion
length $\sigma$ (particle diameter), related with the excluded volume.\cite{Royall_2013,Royall_2007}
Free energy and pressure of few HS colloids confined in a spherical
pore were studied recently by molecular dynamics and theoretical approaches.\cite{Urrutia_2014a,Urrutia_2012,Urrutia_2011_b}
In several cases, it is necessary to complement this hard-core repulsion
with an attractive, short-range interaction, to properly describe
the potential between colloidal particles. For this purpose, the square-well
(SW) potential has been utilized.\cite{Acedo_2001} Systems of particles
that interact through SW potential have been investigated extensively,
exploiting the fact that its simplicity enables to study both, in
bulk and in confinement, not only through Monte Carlo and molecular
dynamics simulations, but also with analytic approaches. SW system
is the simplest interaction model that includes a repulsive core and
an attractive well with tunable range, giving rise to phase transitions
and coexistence regions. The thermodynamic properties, phases coexistence,\cite{Li_2014,RiveraTorres_2013,EspindolaHeredia_2009,Vortler_2008,LopezRendon_2006,Kiselev_2006,Liu_2005}
structure,\cite{Mehrdad_2011} crystallization, glassy behavior\cite{Hartskeerl2009}
and percolation phenomena\cite{Neitsch_2013} of short-range SW fluids
were extensively studied in bulk. Properties of inhomogeneous SW systems
at free interphases and in confinement, were also studied.\cite{ArmasPerez_2013,Neitsch_2013,Huang_2010,Jana_2009,Zhang_2006}
Molecular dynamics studies to obtain bulk free energy of short range
SW particles were performed recently.\cite{RiveraTorres_2013,Pagan_2005}
Pressure on the wall and structural properties were studied for two
SW particles in a spherical cavity.\cite{Urrutia_2011} Liquid-vapor
coexistence of SW fluid confined in cylindrical pores\cite{Reyes_2012}
and short range SW potential in the context of effective interactions
among proteins\cite{Duda_2009} were also studied.

Adding long, flexible, non-adsorbing polymer chains to a colloidal
suspension causes changes on the phase behavior of the colloidal system.
We shall focus on the rather simple Asakura-Oosawa model (AO) for
colloid-polymer mixtures.\cite{Asakura_1954,Asakura_1958} In this
model based on HS-type interactions, the colloids are taken as hard
spheres with diameter $\sigma$. The polymers with diameter $\sigma_{p}$
are excluded by a center of mass distance of $\frac{\sigma+\sigma_{p}}{2}$
from the colloids. However, polymers are treated as non-interacting
particles that can overlap. This kind of interactions leads to an
effective attractive two-body potential between colloids, due to an
unbalanced osmotic pressure arising from depletion.\cite{israelachvili_11}
Although, for a small enough diameters ratio $\sigma_{p}/\sigma$
the two-body effective potential has a short-range attractive well,
and the three- and many-body potentials are null.\cite{Vrij_1976}
In addition, when $\sigma_{p}/\sigma$ is small the main features
of the colloid-colloid effective pair potential are similar to the
simpler SW potential. Studies of the AO model with different values
of diameter ratio have shown interesting properties in bulk, as glassy
states and demixing,\cite{Germain_2007,LopezdeHaro_2015} as well
as in confinement.\cite{Binder_2014,Winkler_2013,Statt_2012}

In this work, we study thoroughly highly confined colloids in spherical
pores, that show different demeanor as compared to their bulk counterparts.
The confining cavity (thought as a nanopore) breaks the translational
symmetry of a system and causes the appearance of spatial variations.
Even farther from bulk case, our aim is to work with low number of
particles and a system size comparable to its constituents elements.
This implies that we are far away from thermodynamic limit. Nevertheless
we work in the frame of statistical mechanics. In addition, low $N$
systems allow for analytic solutions at some degree, thus letting
us to make a direct comparison between theory and simulation results.
On this line, we study the system of few short-range colloids in a
hard-wall spherical pore. The pure colloidal suspension is studied
by event-driven molecular dynamics simulations using the SW model
in a wide range of number densities and relevant temperatures. We
also map the simulated system to a colloid-polymer mixture in a spherical
pore. In this case we adopt the AO model and consider the case of
small ratio $\sigma_{p}/\sigma$. We connect the short-range SW potential
and the AO model by adopting the effective short-range colloid-colloid
pair potential through an extended corresponding-states law, following
that of Noro and Frenkel.\cite{Noro_2000,ValadezP_2012}

The paper is organized as follows, in Sec. \ref{sec:Model} we provide
details of the interaction model and the statistical mechanics theoretical
grounds for both few body systems: colloidal suspensions and colloid-polymer
mixtures in pores. In Sec. \ref{sec:Simulation} we present the simulation
technique and the way in which we obtain canonical ensemble simulations
at constant temperature. Sec. \ref{sec:Results} is devoted to present
the density profiles, pair correlation functions, thermodynamics properties
and phase diagrams of systems of 2, 3, 4, 5 and 6 colloidal particles
in a spherical cavity. There, we analyze the results for the simulated
SW system and also for the equivalent AO system, whenever possible.
We present a final discussion and conclusions in Section \ref{sec:End}.

\section{Theoretical background\label{sec:Model}}

We study a system of $N$ colloidal particles of diameter $\sigma$
confined in a spherical cavity of radius $R_{o}$, at constant temperature
$T$. The particles interact with the cavity through a hard wall potential
which prevents them from escaping to the outside. Thus, the effective
radius of the cavity is $R_{\textrm{eff}}\equiv R_{o}-\sigma/2$,
which represents the maximum possible distance between the center
of the cavity and the center of each colloid. The temperature of the
fluid is determined by the wall temperature $T$, which is fixed.
We adopt the effective volume $V=4\pi R_{\textrm{eff}}/3$ to measure
the size of the available space for particles and consistently define
the mean number density $\rho=N/V$.

Given that we deal with a small number of colloids, the ensembles
equivalence does not apply. The statistical mechanical and thermodynamic
properties of the system with constant $N$ and $T$ is obtained from
its canonical partition function (CPF), $Q_{N}$. One actually works
with the configuration integral (CI), given that kinetic degrees of
freedom integrate trivially. Thus, the CPF reads
\begin{equation}
Q_{N}=\frac{1}{N!}\Lambda^{-3N}Z_{N}\:.\label{eq:CPF}
\end{equation}
Here $\Lambda$ is the thermal de Broglie wavelength and $Z_{N}$
is the CI of the system. For pair interacting particles $Z_{N}=\int_{V}\prod_{<jk>}e_{jk}\, d\mathbf{r}^{N}$,
where the Boltzmann factor for the $j,k$-pair is $e_{jk}=\exp\left[-\beta\phi\left(r_{jk}\right)\right]$,
$r_{jk}$ is the distance between both particles, $\phi$ is the pair
potential and the inverse temperature is defined as $\beta=1/kT$,
with $k$ the Boltzmann constant. For a system in stationary conditions
with fixed $N$ and $T$ the Helmholtz free energy ($F$) reads 
\begin{equation}
F=U-TS=-\beta^{-1}\ln Q_{N}\:,\label{eq:FlogQ}
\end{equation}
where $U$ is the system energy and $S$ its entropy. The reversible
work done at constant temperature, to change the cavity radius between
states $a$ and $b$ is
\begin{equation}
F_{b}-F_{a}=-\intop_{a}^{b}P_{w}dV\:,\label{eq:DiffF}
\end{equation}
with $dV=A\, dR_{\textrm{eff}}$. Here, $P_{w}$ is the pressure on
the spherical wall which is an EOS of the system. The derivative of
Eq. (\ref{eq:DiffF}) at constant $T$ gives the pressure on the wall
through
\begin{equation}
P_{w}=-A^{-1}dF/dR_{\textrm{eff}}\:,\label{eq:PwThermo}
\end{equation}
which meets the exact relation known as contact theorem\cite{Hansen2006,Blokhuis_2007}
\begin{equation}
\beta P_{w}=\rho\left(R_{\textrm{eff}}\right)\:.\label{eq:rhocontact}
\end{equation}
In this ideal gas-like relation, $\rho\left(R_{\textrm{eff}}\right)$
is the value that takes the density profile at contact with the wall.
This extended version of the contact theorem for planar walls applies
to curved walls of constant curvature (spheres and cylinders), for
both open and closed systems. A complete discussion of the presented
statistical mechanical approach for few-body confined system that
includes other properties, such as the energy, may be found in Refs.
\cite{Urrutia_2011,PaganiniTL_2014}.

Here we consider the confined colloids as particles that interact
through the square well potential: 
\begin{equation}
\phi_{\textrm{SW}}\bigl(r\bigr)=\begin{cases}
\infty & \textrm{if }0<r<\sigma,\\
-\varepsilon & \textrm{if }\sigma<r<\left(1+\lambda\right)\sigma,\\
0 & \textrm{if }r>\left(1+\lambda\right)\sigma,
\end{cases}\label{eq:phiSW}
\end{equation}
where $\varepsilon>0$. In this work, we study the short-range SW
system with $\lambda=0.1$. The bulk and interfacial properties of
short-range SW fluids have been studied elsewhere.\cite{ValadezP_2012,Neitsch_2013,LopezRendon_2006}
For reference, we note that the bulk SW system with $\lambda=0.1$
has a metastable fluid-vapor transition with critical temperature
$T=0.47\,\varepsilon/k$ and density $\rho=0.47\,\sigma^{-3}$.\cite{ValadezP_2012,Liu_2005}

\subsection*{Statistical mechanics of the semi-grandcanonical confined AO system\label{sub:StatMechAO} }

In the AO model a particular colloid-polymer mixture is characterized
by the ratio $q=\frac{\sigma_{p}}{\sigma}$. The diameter of the polymer
coil is given by $\sigma_{p}=2R_{g}$ with $R_{g}$ the radius of
gyration of the polymer. The effect of temperature on $R_{g}$ was
studied in Ref.\cite{Taylor_2012}. In the AO model the temperature
does not play any relevant role and thus we consider $T$ as fixed
to fix $\sigma_{p}$. We consider the mixture of $N$ colloidal particles
and polymers at chemical potential $\mu_{p}$, at a given temperature.
The spherically confined system of colloids is such that the polymers,
which are much smaller than colloidal particles, can freely pass through
the semipermeable wall that only constrains the colloids into the
pore. The colloid-polymer AO mixture is an inhomogeneous system that
can be analyzed in the semi-grand canonical ensemble. Its partition
function is
\begin{equation}
\Xi_{m}=\frac{\Lambda^{-3N}}{N!}\sum_{N_{p}}\frac{z_{p}^{N_{p}}}{N_{p}!}Z_{N,N_{p}}\:,\label{eq:SGCPF}
\end{equation}
with $z_{p}=\varLambda_{p}^{-3}e^{\beta\mu_{p}}$, $\varLambda_{p}$
the thermal de Broglie length of the polymer and $\varLambda$ the
same magnitude for the colloidal particle. $Z_{N,N_{p}}$ is the CI
of the mixture with the $N$ colloids, whose centers are constrained
to the pore with volume $V=4\pi R_{\textrm{eff}}^{3}/3$, and the
$N_{p}$ polymers in the larger volume $V_{p}$. In the Appendix \ref{sec:Appendix}
it is shown that Eq. (\ref{eq:SGCPF}) transforms to 
\begin{equation}
\Xi_{m}=\Xi_{p}^{h}\frac{\Lambda^{-3N}}{N!}e^{-\bigl(\rho_{p}v_{exc}N\bigr)}Z_{N}^{(\textrm{AO})}\:,\label{eq:SGCPF2}
\end{equation}
where $\rho_{p}$ is the mean number density of the pure (homogeneous)
polymer system and $Z_{N}^{(\textrm{AO})}$ is the CI of $N$ confined
colloids interacting trough the effective pair-potential $\phi_{\textrm{AO}}$.
$v_{exc}$ stands for the excluded volume defined in the Appendix
\ref{sec:Appendix}. Given two colloids at a distance $r$ apart,
$\phi_{\textrm{AO}}(r)$ is infinity for $r<\sigma$ and it is zero
for $r>\sigma(1+q)$. In the attractive well region $\sigma<r<\sigma(1+q)$
it is
\begin{equation}
\beta\phi_{\textrm{AO}}\bigl(x\bigr)=\!-\eta_{p}\left(1+q^{-1}\right)^{3}\!\biggl[1-\frac{3x}{2\left(1+q\right)}+\frac{x^{3}}{2\left(1+q\right)^{3}}\biggr],\label{eq:potAO}
\end{equation}
where $x=r/\sigma$ and $\eta_{p}=\rho_{p}\left(\pi\sigma_{p}^{3}/6\right)$
is the packing fraction of the polymers (note that $\eta_{p}$ takes
any positive value). The expression in Eq. (\ref{eq:potAO}) is minus
$\rho_{p}$ times the volume of intersection of two spheres of radius
$\sigma\left(1+q\right)/2$ whose centers are at a distance $r$.

For the thermodynamic analysis of the AO model we use as reference
the pure polymer system (an ideal-gas) with the grand-free energy
given by $\Omega_{p}^{h}=U_{p}^{h}-TS_{p}^{h}-\mu_{p}N_{p}^{h}$ (with
$N_{p}^{h}=\rho_{p}V_{p}$). The semi-grand free energy of the mixture
is $\Omega_{m}=U_{m}-TS_{m}-\mu_{p}N_{p}$ and given that the system
is athermal its energy is purely kinetic $U_{m}=\frac{3}{2}kTN+\frac{3}{2}kTN_{p}$.
We define
\begin{eqnarray}
F_{\textrm{AO}} & = & \Omega_{m}-\Omega_{p}^{h}\:,\nonumber \\
 & = & U_{c}-T\left(S_{m}-S_{p}^{h}\right)+\left(3kT/2-\mu_{p}\right)\Delta N_{p}\,,\label{eq:FAO}
\end{eqnarray}
with $U_{c}=\frac{3}{2}kTN$, $\Delta N_{p}=N_{p}-N_{p}^{h}$ and
$\Omega=-\beta\ln\Xi$. Here, $F_{\textrm{AO}}$ is the free energy
of the confined colloids in the polymer solution as an excess over
the pure polymer system. The pressure exerted by the colloids on the
wall, the osmotic pressure, is 
\begin{equation}
P_{w}=-A^{-1}dF_{\textrm{AO}}/dR_{\textrm{eff}}\:,\label{eq:PwAO}
\end{equation}
and relates with the colloids density distribution $\rho\left(\mathbf{r}\right)$
through the contact theorem 
\begin{equation}
\beta P_{w}=\rho\left(R_{\textrm{eff}}\right)\:.\label{eq:rhocontactAO}
\end{equation}
Eqs. (\ref{eq:PwAO}, \ref{eq:rhocontactAO}) are essentially the
same that Eqs. (\ref{eq:PwThermo}, \ref{eq:rhocontact}) but the
meaning of each magnitude corresponds to different systems.

\subsection*{Extended law of corresponding states\label{sub:Noro-Frenkel}}

The short range SW potential and its capability to describe any short
range potential (\emph{universality}) was proposed by Noro and Frenkel
in his extended version of the corresponding state law.\cite{Noro_2000}
It is based on a mapping between different systems using three parameters:
the effective hard core diameter, the well depth and the adimensional
second virial coefficient. The later was proposed as a measure for
the range of the attractive part of the potential. The scheme was
used previously for studying the critical properties of the liquid-vapor
transition for interaction models including Lennard-Jones and Hard-Yukawa.\cite{Noro_2000,Germain_2010_b,Gnan_2012}.
Recently, it was employed to analyze the behavior of proteins in water
solutions.\cite{ValadezP_2012}

It is important to note that we apply the law of corresponding states
to analyze confined systems composed by few particles. This is very
unusual and thus we made some checks to validate the overall approach
that will be shown in Sec. \ref{sec:Results}. Our application of
the law for hard-core systems is based on the use of two natural scales:
the hard core diameter for the length scale and the depth of the attractive
well for the temperature scale. The reduced second virial coefficient
is given by $B=-\frac{1}{2b_{2}}\int\left[\exp\left(-\beta\phi\bigl(r\bigr)\right)-1\right]\, d\mathbf{r}$
with $b_{2}=2\pi\sigma^{3}/3$. For the SW system, this gives explicitly:
\begin{equation}
B_{\textrm{sw}}=1-\left(e^{1/T^{*}}-1\right)3\lambda\left(1+\lambda+\lambda^{2}/3\right)\:,\label{eq:Bsw}
\end{equation}
where we have introduced the adimensional temperature $T^{*}=Tk/\varepsilon$
(its inverse is $\beta^{*}=1/T^{*}$). In the limit of a very narrow
well, the SW potential reaches the Baxter's sticky spheres limit with
\[
B=1-1/4\tau\:.
\]
Here $\tau$ (that grows monotonically with $T$) plays the role of
temperature.\cite{HansenGoos_2012,Charbonneau_2007} In the AO model
the effective colloid-colloid second virial coefficient $B_{\textrm{AO}}$
is not analytically integrable. To relate $B_{\textrm{sw}}$ with
$B_{\textrm{AO}}$, we link the width of the wells by $q=2\lambda$.
This mimics the fact that $\beta\phi_{\textrm{AO}}\bigl(x\bigr)$
is deeper near the hard-core of the particle. For the AO system with
corresponding well-range $q=2\lambda$, the relation $B_{\textrm{AO}}\left(\eta_{p},q=2\lambda\right)=B_{\textrm{sw}}\left(T^{*},\lambda\right)$
fixes the pair of equivalent states $T^{*}\longleftrightarrow\eta$.
\begin{table}
\centering{}%
\begin{tabular}{|c|c|c|c|}
\hline 
$q=2\lambda$ & $c_{1}$ & $c_{2}$ & $c_{3}$\tabularnewline
\hline 
\hline 
$0.2$ & $0.886964$ & $0.22724$ & $0.015641$\tabularnewline
\hline 
$0.15$ & $0.884047$ & $0.29956$ & $0.013871$\tabularnewline
\hline 
$0.1$ & $0.876335$ & $0.44847$ & $0.012614$\tabularnewline
\hline 
$0.05$ & $0.847353$ & $0.92875$ & $0.013498$\tabularnewline
\hline 
\end{tabular}\protect\caption{Fitting parameters for the SW-AO mapping. They relate $T^{*}$ of
the SW system to the value of the packing fraction $\eta_{p}$ (see
Section \ref{sub:StatMechAO}) in the corresponding state of the AO
system.\label{tab:FitingParam}}
\end{table}
 This mapping allows the numerical evaluation of $B_{\textrm{AO}}$
and the fit of a nearly linear relation between $1/T^{*}$ and $\eta$
for several values of $\lambda$. The coefficients obtained for the
fitted function 
\begin{equation}
\frac{1}{T^{*}}=\frac{2}{3}\left(1+1.5q^{-1}\right)\sum_{i=1}^{3}c_{i}h_{i}\:,\label{eq:AO-SW-map}
\end{equation}
with $h_{1}=\eta_{p}$, $h_{2}=\eta_{p}^{2}$ and $h_{3}=\eta_{p}/(0.1+\eta_{p})$,
are shown in Table \ref{tab:FitingParam}. 
\begin{table}
\begin{centering}
\begin{tabular}{|c|c|c|}
\hline 
$\tau$ & $T^{*}\,\left(\textrm{SW, }\lambda=0.1\right)$ & $\eta_{p}\,\left(\textrm{AO, }q=0.2\right)$\tabularnewline
\hline 
\hline 
$0.0051$ & $0.2$ & $0.81071$\tabularnewline
\hline 
$0.0675$ & $0.4$ & $0.43466$\tabularnewline
\hline 
$0.1758$ & $0.6$ & $0.29597$\tabularnewline
\hline 
$0.3033$ & $0.8$ & $0.22369$\tabularnewline
\hline 
$0.4395$ & $1.0$ & $0.17939$\tabularnewline
\hline 
$0.5805$ & $1.2$ & $0.14950$\tabularnewline
\hline 
\end{tabular}
\par\end{centering}

\protect\caption{Law of corresponding states for short range potentials of SW and AO
types. First column presents the sticky-sphere temperature parameter
$\tau$. Second and third columns show the temperature for the SW
particles and the corresponding packing fraction of polymer for the
AO model, respectively.\label{tab:The-law-of}}
\end{table}
The case $\lambda=0.1$ and $q=0.2$ is presented in Table \ref{tab:The-law-of},
where corresponding values of sticky temperature, adimensional temperature
of the SW system and polymer packing fraction are shown. 
\begin{figure}
\begin{centering}
\includegraphics[width=7cm]{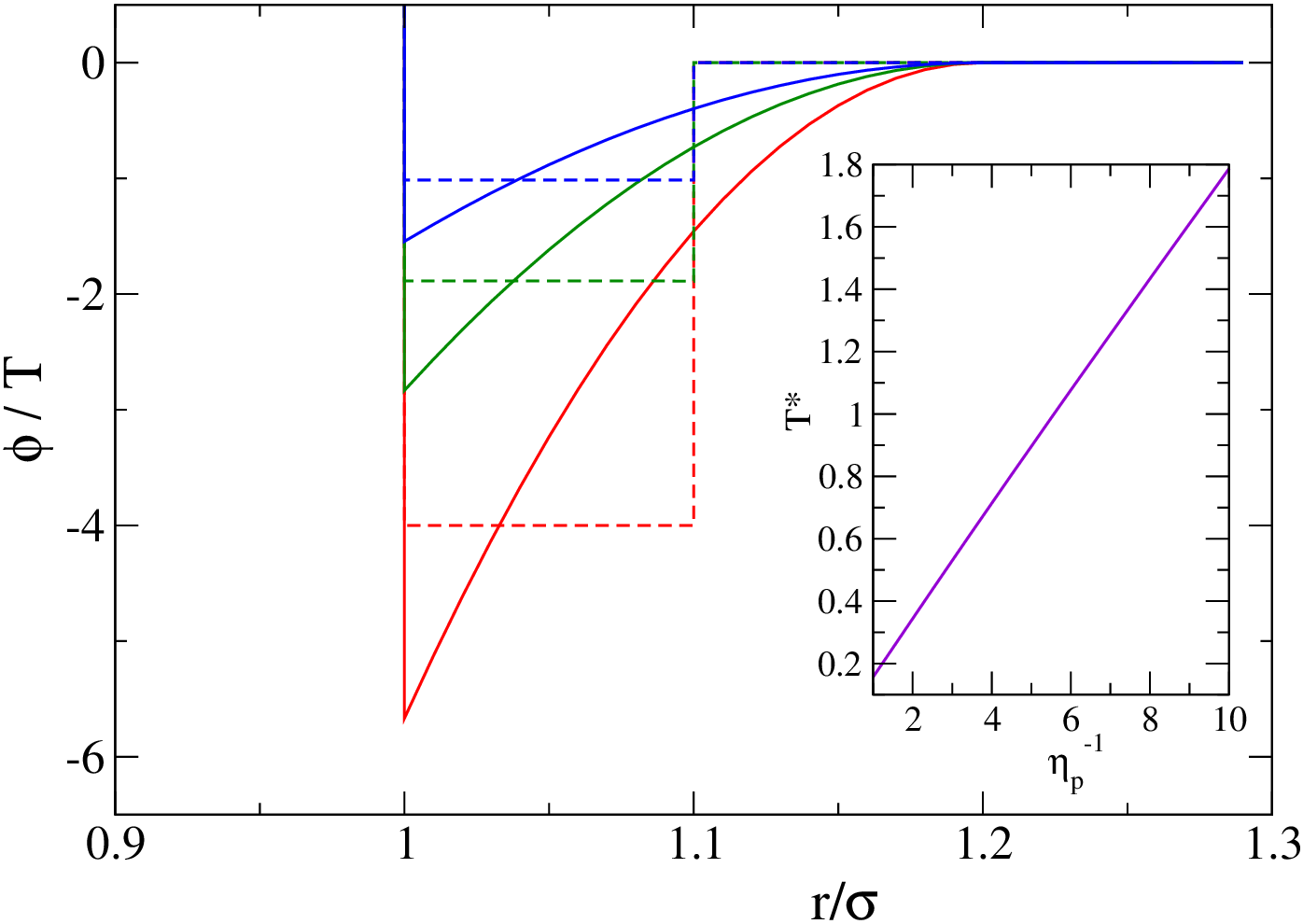}
\par\end{centering}

\protect\caption{Corresponding states between SW potential (dashed lines) and AO effective
potential (full lines) for $\lambda=0.1$ and $q=0.2$. From bottom
to top, the inverse of the packing fraction (temperature) for AO (SW)
curves are $\eta_{p}^{-1}=1.5$, $3$ and $5.5$ ($T^{*}=0.25$, $0.53$
and $0.985$). The inset presents the nearly linear relation between
$T^{*}$ and $\eta_{p}$ used to map the SW system to the AO model.\label{fig:NoroFrenkel-like}}
\end{figure}
In Fig. \ref{fig:NoroFrenkel-like} we present the SW potential and
the effective AO potential that yields corresponding states for several
temperatures and polymer packing fraction ($\lambda=0.1$ and $q=0.2$).
The inset shows the simple relation between the temperature of the
SW particles and the corresponding packing fraction of polymers for
the AO colloid-polymer mixture expressed in Eq. (\ref{eq:AO-SW-map}).
In Sec. \ref{sec:Results}, the overall approach will be used to analyze
the spherically confined system of few colloids in a colloid-polymer
mixture.

\section{Simulation method\label{sec:Simulation}}

The statistical mechanical equivalence between different ensembles
does not apply to few-body systems. Therefore, simulation and statistical
mechanical approaches should correspond to the same physical constraints,
to ensure comparable results. In this work, we focus on a system at
constant temperature, fixed volume and number of particles. It corresponds
to a canonical ensemble, and thus one assumes a Maxwell-Boltzmann
velocity distribution. Accordingly, the simulations have to include
a thermostating mechanism that ensures constant temperature and Maxwell-Boltzmann
velocity distribution.

Molecular dynamics simulations of few SW confined in a spherical cavity
were performed with a standard event-driven algorithm (EDMD).\cite{Allen_and_Tildesley}
Constant temperature was achieved by using a thermal-wall thermostat,
which changes the velocity of the particle colliding with the wall
by means of a velocity distribution compatible with canonical ensemble
for a given temperature $T$. Thermal walls show certain features
different from those thermostats that act over the entire system volume
that were discussed in detail in a previous work.\cite{Urrutia_2014a}

For the particle interactions, we take into account different types
of events. Namely, particle-particle collision and particle-wall collision.
Among particle-particle collisions there is a further division between
core and field events, each one related with a discontinuous step
in $\phi_{\textrm{SW}}\bigl(r\bigr)$.%
{} The usual EDMD algorithm was used, in which the particle moves with
rectilinear and constant-velocity dynamics, between particle collisions.\cite{Allen_and_Tildesley,LopezRendon_2006}
In order to discriminate particle-particle collisions we used the
logical structure of Alder and Wainwright.\cite{Alder_1959}

The time to collision of particle $i$ with particle $j$ is calculated
as:
\begin{equation}
t_{ij,k}=\frac{-b_{ij}\pm\left[b_{ij}^{2}-v_{ij}^{2}\left(r_{ij}^{2}-\sigma_{k}^{2}\right)\right]^{1/2}}{v_{ij}^{2}}\:,\label{eq:t_ij_edmd}
\end{equation}
where $\mathbf{r}_{ij}\equiv\mathbf{r}_{i}-\mathbf{r}_{j}$ and $\mathbf{v}_{ij}\equiv\mathbf{v}_{i}-\mathbf{v}_{j}$
are the relative positions and velocities of the particle pair, respectively.
Two collision distances $\sigma_{k}$ are considered $\sigma_{1}=\sigma$
and $\sigma_{2}=\sigma\left(1+\lambda\right)$. The parameter $b_{ij}\equiv\mathbf{r}_{ij}\cdot\mathbf{v}_{ij}$
must be negative if the particles are approaching each other and positive
otherwise, and we consider only the positive values of $t_{ij,k}$.
The Eq. (\ref{eq:t_ij_edmd}) is obtained by imposing the condition
$|\mathbf{r}_{ij}+\mathbf{v}_{ij}t_{ij,k}|=\sigma_{k}$ at collision
time $t_{ij,k}$. The $\pm$ appears because both results are possible,
given the right conditions: the ``$-$'' sign applies for particles
approaching from outside regions ($r_{ij}>\sigma_{2}$), and a ``$+$''
sign, on the other hand, is for particles already inside the attractive
well region ($r_{ij}<\sigma_{2}$) and getting outside of it, by receding.

An ordered list of events, with increasing collision times $t_{ij,k}$
is generated. Between collisions, the particles move with $\mathbf{r}_{i}=\mathbf{v}_{i}t$.
Once a collision occurs, the new velocities of the pair of particles
involved in the collision are obtained as:
\begin{eqnarray*}
\mathbf{v}_{i}^{{\rm new}} & = & \mathbf{v}_{i}^{{\rm old}}+\delta\mathbf{v}\:,\\
\mathbf{v}_{j}^{{\rm new}} & = & \mathbf{v}_{j}^{{\rm old}}-\delta\mathbf{v}\:,
\end{eqnarray*}
by momentum and energy conservation $\delta\mathbf{v}$ is easily
calculated. For core collisions
\begin{equation}
\delta\mathbf{v}=\frac{-\mathbf{r}_{ij}b_{ij}}{\sigma^{2}}\:.
\end{equation}

Field cases present different possibilities. In a first event type,
a particle enters the field reducing pair potential energy and consequently
increasing the kinetic energy:
\begin{equation}
\delta\mathbf{v}=\frac{-\mathbf{r}_{ij}}{2\sigma^{2}\left(1+\lambda\right)^{2}}\left[\left(\frac{4\sigma_{2}^{2}\varepsilon}{m}+b_{ij}^{2}\right)^{\frac{1}{2}}+b_{ij}\right]\:.
\end{equation}
Secondly, a particle is leaving the field and the pair has enough
energy to break the bond. Then the potential energy increases requiring
a subtraction from the kinetic energy:
\begin{equation}
\delta\mathbf{v}=\frac{-\mathbf{r}_{ij}}{2\sigma^{2}\left(1+\lambda\right)^{2}}\left[-\left(-\frac{4\sigma_{2}^{2}\varepsilon}{m}+b_{ij}^{2}\right)^{\frac{1}{2}}+b_{ij}\right]\:.
\end{equation}
Finally, the particles separate from each other to leave the field
but there is not enough energy to surpass the well depth. Then a bounce
occurs as if it were a hard collision
\begin{equation}
\delta\mathbf{v}=\frac{-\mathbf{r}_{ij}b_{ij}}{\sigma^{2}\left(1+\lambda\right)^{2}}\:.
\end{equation}

It must be considered in addition, the time at which each particle
collides with the wall $t_{i}^{w}$. This time is calculated by the
condition
\[
|\mathbf{r}_{i}+\mathbf{v}_{i}t_{i}^{w}|=R_{\textrm{eff}}\:,
\]
The nearest next event is chosen as the minimum of the next particle
and wall events: $\min(\min(t_{ij,k}),\min(t_{i}^{w}))$. If the particle-wall
collision is the next event, the system is evolved until the particle
reaches the wall. At this point, the thermal-wall thermostat acts
on the particle by imposing it a new velocity, which is chosen stochastically
from the probability distributions:
\begin{eqnarray}
p_{n}(v_{n}) & = & m\beta|v_{n}|\exp\left(-\beta\frac{1}{2}mv_{n}^{2}\right)\label{eq:thermal_wall}\\
p_{t}(v_{t}) & = & \sqrt{\frac{m\beta}{2\pi}}\exp\left(-\beta\frac{1}{2}mv_{t}^{2}\right)\:,\nonumber 
\end{eqnarray}
here $n$ and $t$ stands for the normal and tangential components
of the velocities, that lie in directions $-\hat{r}$ and $\mathbf{v}^{{\rm old}}-(\mathbf{v}^{{\rm old}}\cdot\hat{r})\hat{r}$,
respectively. The thermal walls described by Eq. (\ref{eq:thermal_wall})
fix the temperature of the system. They were tested for HS confined
both by planar walls and in a spherical pore, and produce a velocity
distribution compatible with that of Maxwell-Boltzmann\cite{Tehver_98,Urrutia_2014a}.
This thermal wall functionality has been extensively tested in a previous
work with focus in confined HS particles\cite{Urrutia_2014a} and
we get the same results for simulations with SW particles. 

Setting an $N$ value between 2 and 6, we swep the $(T,\rho)$ surface.
For each chosen point of that surface we performed 10 simulations
of $3\times10^{6}$ collision events, with a further average of the
results. Temperature range was taken to cover two clearly distinct
behavior regions. At low $T$ particles rack up and form a cluster
that acts as a rigid body, at high values particles dissociate resembling
a HS system. Densities were picked from the very low ``bulk like''
to high values, in the vicinity of close-packing condition.

From simulations we get to measure several quantities attained from
time averages over all systems configurations. The studied structure
and position functions are: the one body density function $\rho(\mathbf{r})$
and the \emph{averaged} pair distribution function $\bar{g}(r)$.\foreignlanguage{spanish}{\cite{Urrutia_11}}
The upper bar is just to distinguish our measured function from the
better known radial distribution function $g(r)$ that is commonly
used to study homogeneous and isotropic systems. The calculated profiles
for $\rho(r)$ and $\bar{g}(r)$ are mean values over a discrete domain,
obtained from a binning of spherical shells during the elapsed simulation
time. The bin length is established by dividing the maximum possible
distance value ($R_{\textrm{eff}}$ for $\rho(r)$ and $2R_{\textrm{eff}}$
for $\bar{g}(r)$) by  the number of desired bins, usually $1200$.

\section{Results\label{sec:Results}}

In this section we present the properties for the confined system
of few short-range SW particles ($\lambda=0.1$) obtained from the
molecular dynamics simulation. They are separated in structural properties,
thermodynamic observables, and phase diagram. Structural behavior
is analyzed in terms of the spatial correlation between particles
from $\bar{g}(r)$ and their spatial distribution in the cavity from
$\rho(\mathbf{r})$. The measured thermodynamic quantities describe
the system as a whole, focusing on magnitudes that are usually utilized
to characterize both bulk and inhomogeneous fluids composed by many-bodies.
The statistical mechanical background was developed in Sec. \ref{sec:Model}
and in Ref. \cite{Urrutia_11}. Phase diagrams should not be understood
on the context of bulk phases. Even, they condense the observed system
behavior in the temperature/density plane. For simplicity all the
magnitudes are presented using natural units, i.e. the unit of length
is $\sigma$, the unit of temperature is $k/\varepsilon$ and the
unit of energy is $\varepsilon$.

Additionally, we discuss the extent to which we expect an accurate
mapping of different structural and thermodynamic properties between
the AO and the SW systems. We analyze a few properties of the equivalent
AO system based on the mapping between $T$ and the packinf fraction
$\eta_{p}$. The presented overall application of the extended law
of corresponding states between AO and SW confined system is tested
at the level of phase diagram.

\subsection{Structural description\label{sub:ResStructure}}

\begin{figure}
\includegraphics[width=0.98\columnwidth]{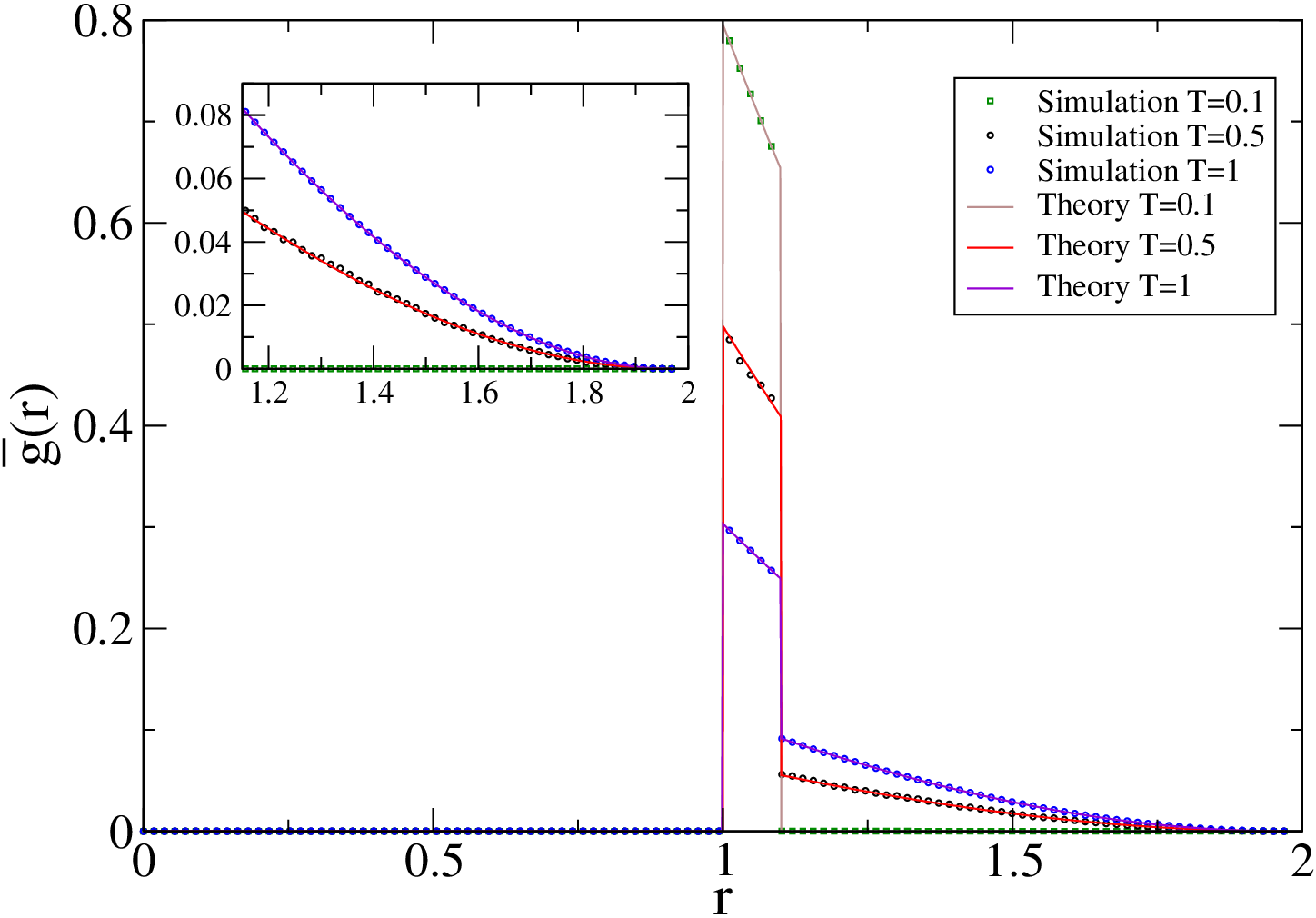}

\protect\caption{\label{fig:pairN2}Pair distribution function curves for $N=2$ and
$\rho=0.5$ for different temperatures. A clear agreement between
theoretical and simulation results is observed. The inset shows same
curves on the range $r\in\left[1.15,1.97\right]$.}
\end{figure}
We present firstly the results for the $N=2$ system, from which exact
theoretical results are available.\foreignlanguage{spanish}{\cite{Urrutia_11}}
This allows to make a direct comparison between exact predictions
and simulation results. By getting a perfect matching, we ensure that
we have solid framework to analyze the few-particle systems of higher
$N$. It is also useful for validating the simulation program and
thermostating procedure. The pair distribution function is shown in
Fig. \ref{fig:pairN2} for different temperatures. A clean superposition
between theory and simulation is easily appreciable. For $N=2$ the
analytic form of $\bar{g}(r)$ is 
\begin{equation}
\bar{g}(r)=Ce_{12}(r)\left[(2R_{\textrm{eff}}-r)^{2}(r+4R_{\textrm{eff}})\right],\label{eq:pairN2}
\end{equation}
with $e_{12}(r)=e^{-\beta\phi(r)}$ and $C=\frac{\pi}{12Z_{2}}$.\cite{Urrutia_11}
First, the null values in the range $0<r<1$ is a trivial effect of
the hard core repulsion. Then in the range $1<r<1.1$ there is always
a main peak. This peak is an expected result, given the shape and
length of the potential: inside the well attractive zone, particles
are more likely to be closer. Another common element of these curves
is the presence of ``tails'', i.e. the smooth monotonically decreasing
segments that are seen for ranges of $r$ beyond the main peak. The
tail ends at $2R_{\textrm{eff}}=1.97$ (cavity diameter) which is
the maximum possible pair distance. Since there is not particle interaction
for pair distance over $1.1$, the tail is proportional to the probability
of finding a pair of hard spheres in spherical confinement at a given
distance. The relation between the main peak and the tail sizes is
driven by the temperature: a steep jump in $e_{12}$ will happen at
low $T$, and $e_{12}$ approaches to unity at high $T$. From a phenomenological
standpoint, at low $T$ the particles lack the energy to escape from
the well, thus forming a permanent short ranged bond. At high $T$,
the kinetic energy is far greater than the well depth, meaning that
the system resembles one of colliding hard cores (HS limit). For higher
$N$, the competition between the structure (peaks) and the tail (no
interaction) is one of the most visible effects when increasing $T$.
The relation between $\bar{g}_{\textrm{sw}}(r)$ and $\bar{g}_{\textrm{AO}}(r)$
was not studied before, at the best of our knowledge. However, based
on our analysis of the case $N=2$, we propose that the corresponding
functions are $\bar{g}_{\textrm{sw}}(r)/e_{12}^{(\textrm{sw})}(r)$
and $\bar{g}_{\textrm{AO}}(r)/e_{12}^{(\textrm{AO})}(r)$, with $r>1$.
This mapping produces small changes in the shape of the main peak
of $\bar{g}_{\textrm{AO}}(r)$ in comparison with $\bar{g}_{\textrm{sw}}(r)$.

\begin{figure}
\includegraphics[width=0.98\columnwidth]{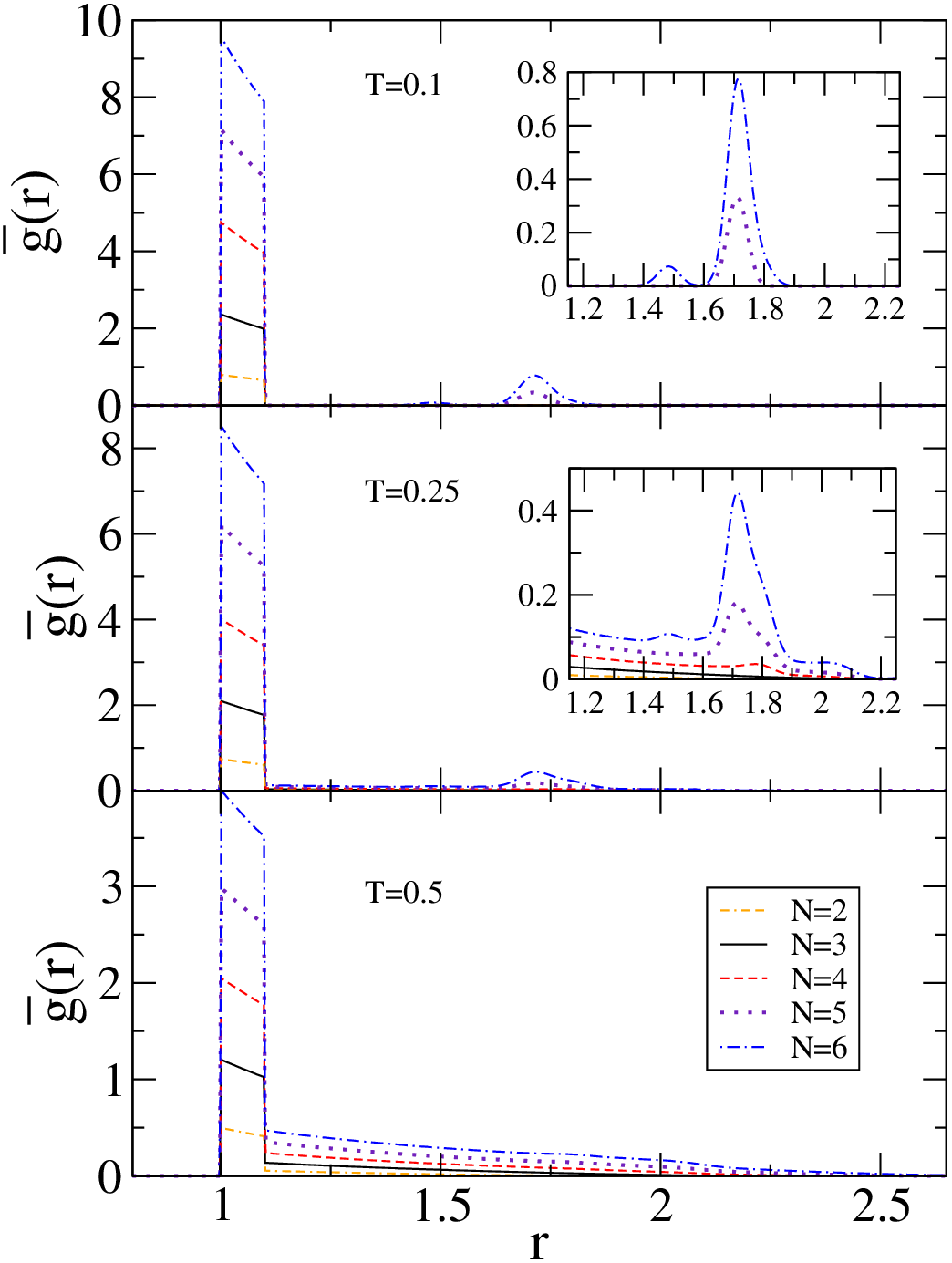}\protect\caption{\label{fig:paircomp1}$\bar{g}(r)$ curves for $N=2$ to $6$ at fixed
number density $\rho=0.5$. Chosen temperatures are (top to bottom
in order) $0.1$, $0.25$ and $0.5$. The insets present the second
neighbour peak in more detail.}
\end{figure}
As can be seen in Fig. \ref{fig:paircomp1}, where it is shown the
pair distribution function for 3 to 6 particles inside the cavity,
adding particles to such a small system will necessarily cause qualitative
changes beyond the two-body analysis. Nonetheless, certain core aspects
remain the same: those that are linked to the potential shape (main
peak and HS limit). Since the integral over the complete space of
$\bar{g}(r)$ is $N(N-1)/2$,\foreignlanguage{spanish}{\cite{Urrutia_11}}
it is expected that lower $N$ (at fixed $\rho$) show overall lower
curves. Also, $R_{\textrm{eff}}=\left(3N/4\pi\rho\right)^{1/3}$ means
that at low $N$ values and fixed number density, cavity size is highly
susceptible to add or subtract a single particle. The shape of $\bar{g}(r)$
for different temperatures can be used to provide qualitative aspects
of the morphology of the clusters. For very low temperatures ($T<0.1$)
the particles form a rigid cluster minimizing the overall system energy.
Every close neighbor ($1<r<1.1$) adds up $-\varepsilon$ to the potential
energy. The bonds are mainly permanent, meaning stable pair distances
leading to clearly discernible peaks. 

For these small systems, it is possible to interpret the low $T$
curves $\bar{g}(r)$ just by considering the clusters shapes. Three
particles form a triangle and four a regular tetrahedron, both of
which have in common that all the particles are first neighbors between
themselves. This means that the whole probability of finding a pair
at certain distance will localize in the first neighbor range of the
pair distribution function (main peak). This isolated peak is shown
for $N=2,\,3$ and $4$ in Fig. \ref{fig:paircomp1}, top panel. 

The shape of $\bar{g}(r)$ for $N=5$ can be understood starting from
the $N=4$ regular tetrahedron and then adding an extra particle on
one of its faces. The result is an hexahedron composed of two regular
tetrahedrons in contact by one face. The resulting structure has three
particles, each one with four first neighbors and two particles with
three first neighbors and one second neighbor. A cluster geometry
with second neighbors gives rise to a non vanishing probability of
finding a pair distance larger than the main peak region. The stable
structure with the second neighbor distance in a constrained region
results in a second peak.

For the case $N=6$ there are two observed cluster geometries. One
of them is the regular octahedron. However, the most frequent geometry
observed in the simulations is an irregular octahedron. This irregular
polyhedron has a typical path of formation starting from a hexahedral
cluster of five particles to which the remaining particle adds over
one of its faces. For extremely low temperatures ($T\ll0.1$), once
the particles form their bonds, they will stay bonded permanently.
For a softer cluster ($T\lesssim0.1$), single bonds have a slight
chance of breaking and a rearranging of the cluster structure can
take place. The secondary peak of $\bar{g}(r)$ is sensitive to these
different structures as shown in the inset of the top panel in Fig.
\ref{fig:paircomp1}. There are two secondary peaks both related with
second neighbor distance: the larger one represents the second neighbors
in the irregular octahedron while the smaller one is characteristic
of the regular body. The significant height difference shows that
the irregular cluster is more frequent over time. We point out also
that both geometries have the same number of first neighbors being
thus isoenergetic. Therefore, from a statistical mechanics point of
view, the only factor that can lead to the prevalence of one geometry
over the other is strictly coming from entropic contributions. The
irregular cluster presents lower symmetry and higher entropy.

Increasing the temperature leads to the break up of multiple bonds,
which results in flexible or plastic clusters. For intermediate $T$
ranges ($0.1<T<0.25$), the particles remain constantly linked, but
now they are not tightly bound. Certain bonds are likely to break,
enabling the particles to displace inside the cluster. The inset of
Fig. \ref{fig:paircomp1} for $T=0.25$, shows that the isolated peaks
are surrounded by non-vanishing values. For example, the triangle
for $N=3$ breaks one of its pair bonds in such a way that it opens
up and stretches like a chain. For higher $N$, is essentially the
same. Additional translational freedom leads to possible deformations
into wider shapes than the original rigid body. The weaker the structure
is, the lower the main and second peak become, favoring pair distance
probability on the surrounding areas.

For higher temperatures, ($T>0.3$) soft cluster starts to dissociate
and single particles are free to have any distance from the cluster,
inside the cavity limits. Any kind of stable structure fades out and
only instantaneous pairs endure. As can be seen in the bottom panel
of Fig. \ref{fig:paircomp1}, the tail engulfs any close range structure
and the main peak reduces around half of its height, compared to the
$T=0.25$ panel. Higher temperatures do not add any qualitative variation:
the main peak will decrease until it becomes part of the tail, in
the HS limit.

\begin{figure}
\includegraphics[width=7cm]{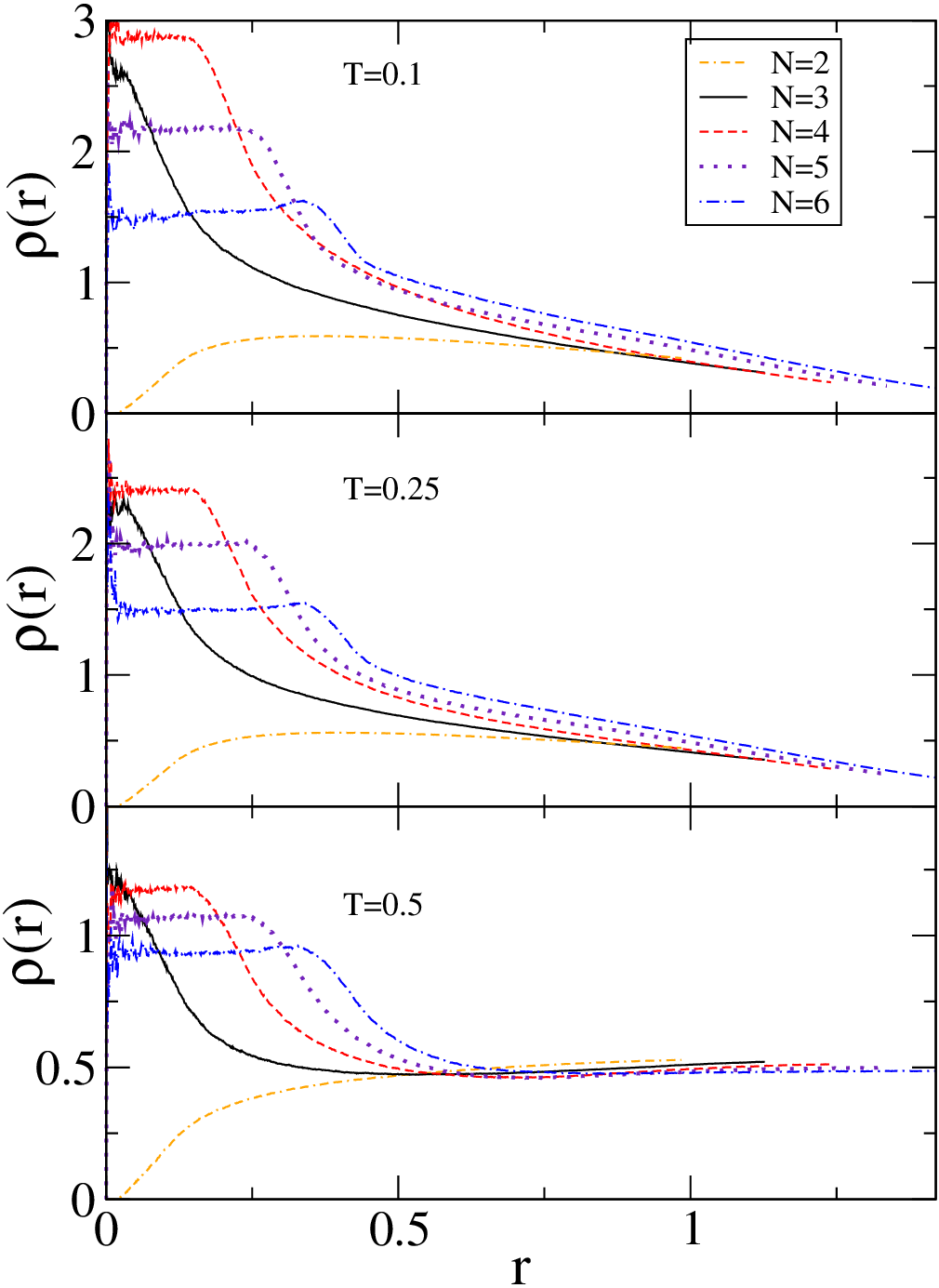}

\protect\caption{\label{fig:denscomp1}Density profiles for $N=2$ to $6$ at fixed
number density $\rho=0.5$. The chosen temperatures are (top to bottom
in order) $0.1$, $0.25$ and $0.5$.}
\end{figure}
Density profiles, shown in Fig. \ref{fig:denscomp1}, present very
explicitly the inhomegeneity of the system. Unlike commonly studied
bulk systems, there are significant local spatial variations of the
one body density function. Note that $R_{\textrm{eff}}$ varies with
$N$, in Fig. \ref{fig:denscomp1}. It takes values from $R_{\textrm{eff}}=1$
for two particles to $R_{\textrm{eff}}=1.9$ for six particles. For
the case $N=2$ the density vanishes in the center because if one
particle in placed at $r\approx0$, the available volume for the other
one becomes very small. For density values that define effective radii
higher than the particle diameter, i.e. $N=3$ to $6$, all the profiles
have similar form, independently of particle number. The profiles
have two distinct regions. An approximately constant density on the
center of the cavity, that is cut at the vicinity of the wall ($R_{\textrm{eff}}-r<1$)
and the \textquotedbl{}interfacial\textquotedbl{} region closer to
the wall. The extension of the plateau depends on the relation of
cavity size and particle diameter, as observed from the different
sizes at equal number density in Fig. \ref{fig:denscomp1}. For low
temperatures the system may be treated as a single nearly-rigid body,
free to translate and rotate in the central region. When the cluster
gets closer to the wall, some possible cluster orientations are restricted,
leading to a reduction in rotational entropy. Consequently, it is
more likely to find the cluster in the central region. Higher temperatures
soften the particles' bonds progressively, causing a reduction of
the disparity between the plateau and the region close to the wall.
Once dissociation becomes dominant for high temperatures, depletion
arises and the wall starts to have a more intense effective attraction.
Further increase in the temperature makes the wall attraction higher,
while the particle correlation disappears. At the HS limit the highest
probability of finding a particle is at the wall. This was also observed
in a system of pure HS particles in spherical confinement.\cite{Urrutia_2014a}
For the AO system we expect similar density profiles to those of SW
for equivalent temperature and packing fraction. The temperatures
$T=0.1$, $0.25$ and $0.5$ correspond to (top to bottom panels in
Fig. \ref{fig:denscomp1}) polymers inverse packing fraction $\eta_{p}^{-1}=0.694$,
$1.50$ and $2.84$, respectively. As we will see, the case $T=0.1$,
may be a too small temperature to use the extended law of corresponding
states.

\begin{figure}
\includegraphics[width=7cm]{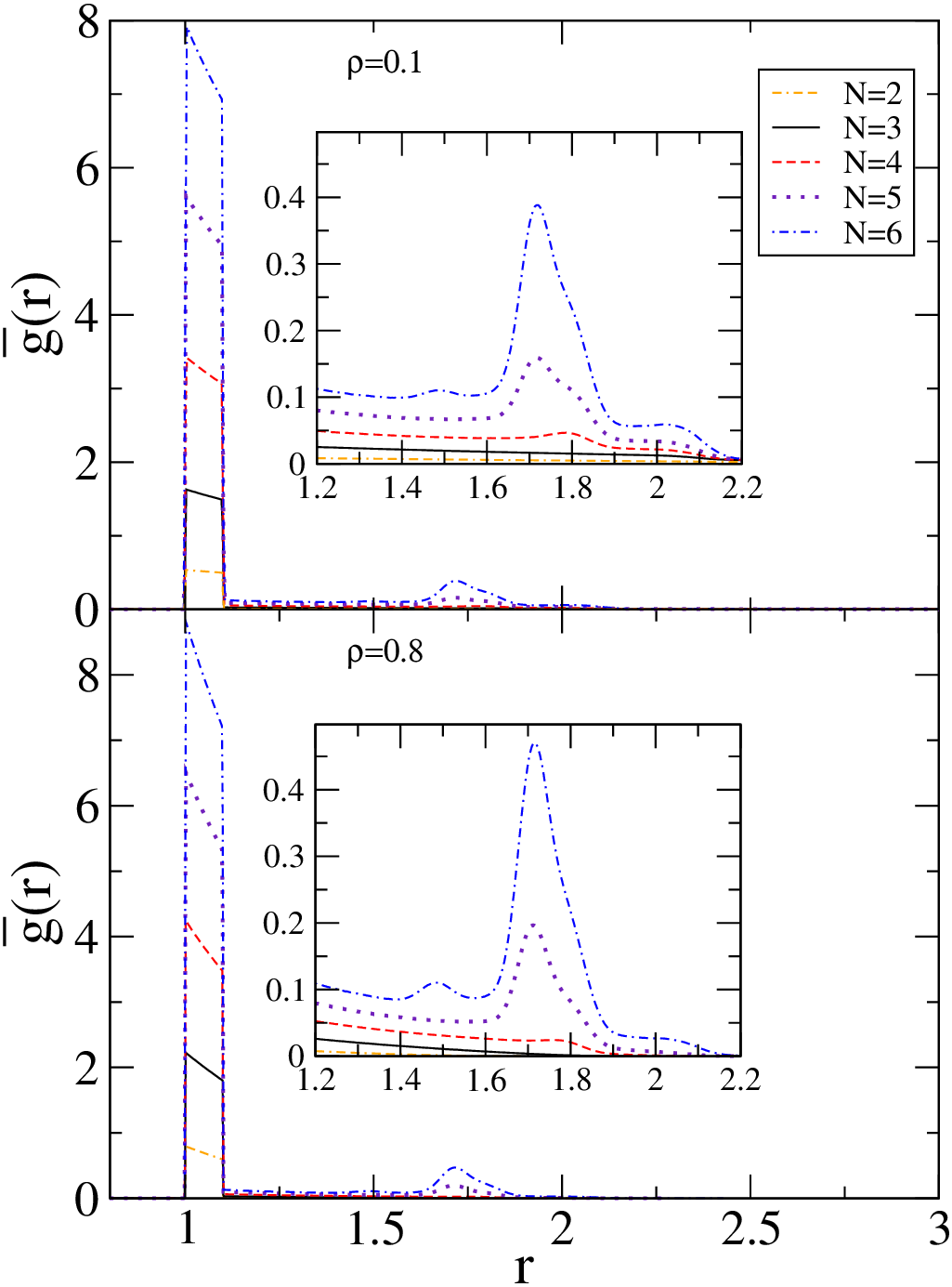}\protect\caption{\label{fig:paircomp2}$\bar{g}(r)$ curves for $N=2$ to $6$ at fixed
temperature $T=0.25$. Number densities are $0.1$ (Top) and $0.8$
(Bottom). The insets present the second neighbour peak in more detail.}
\end{figure}
The structure dependency on $\rho$ is far more subtle than on $T$.
In Fig. \ref{fig:paircomp2} curves of $\bar{g}(r)$ at $T=0.25$
and for two different densities are shown. One observes that the shape
of the peaks remain practically unaltered. Global values of the curve
raise for higher number densities, because the normalization is the
same in a smaller cavity size. We have shown the case $T=0.25$, as
an example but we observed the same general picture for other values
of $T$ which are not presented here.

\begin{figure}
\includegraphics[width=7cm]{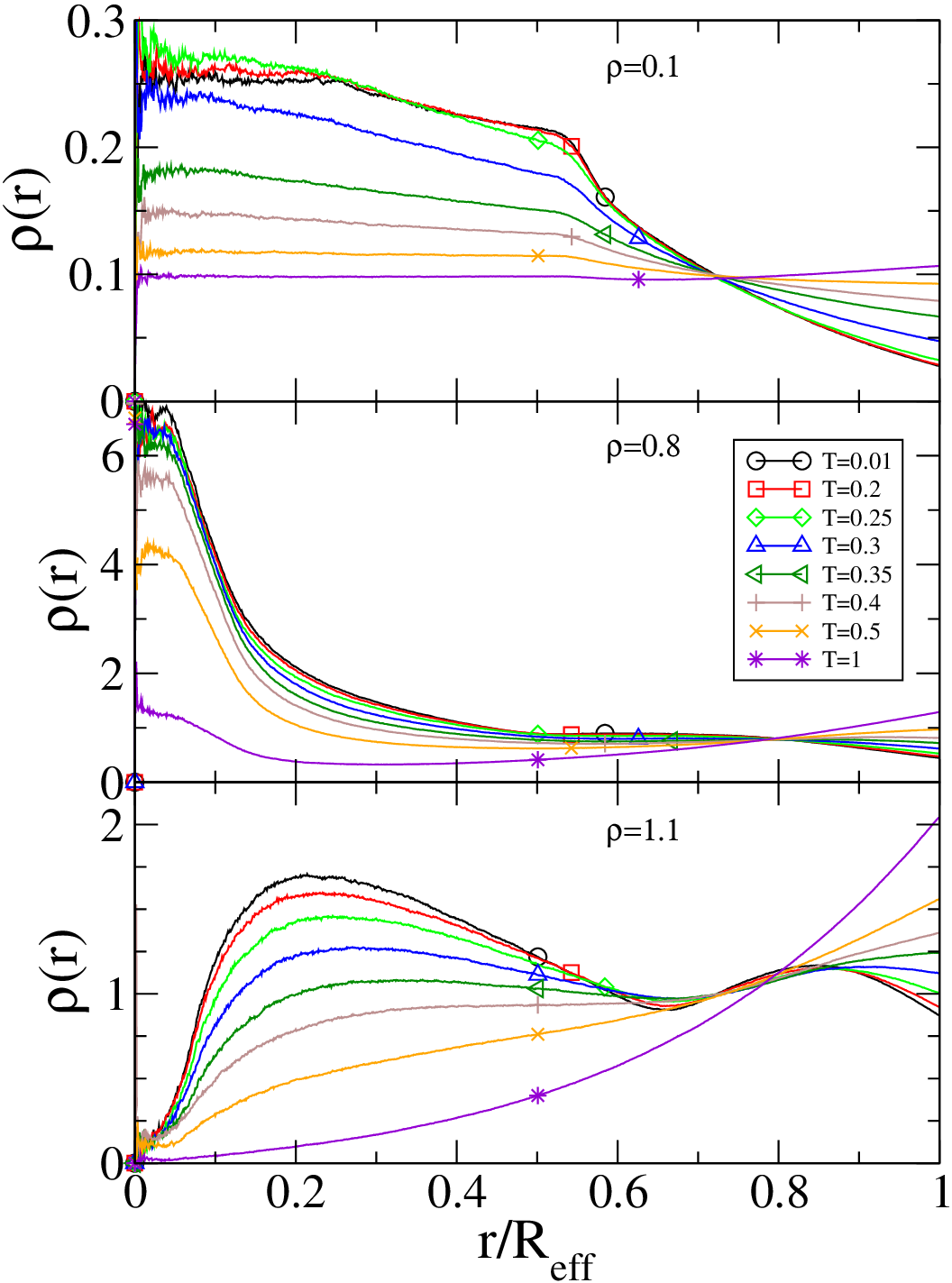}

\protect\caption{\label{fig:densN5}Density profiles for $N=5$ at different temperatures.
Number densities are $0.1$ (Top), $0.8$ (Center) and $1.1$ (Bottom).}
\end{figure}
From here on,  we select the case $N=5$ to give more detailed analysis.
This case is the only one that presents a second peak but does not
have multiple stable geometries, as the case $N=6$. At low temperatures
it is observed a clearly defined structure, while retaining little
longer range order. The $N=5$ results can be extrapolated to the
other few-particle systems. Density profiles are shown in Fig. \ref{fig:densN5}
for a wide range of temperatures and number densities. As already
pointed out, at low and intermediate densities there exist a plateau
and an interfacial region close to the wall. Increasing the temperature
leads to an overall probability density favoring position closer to
the wall, as a result of relatively stronger depletion attraction.\cite{Urrutia_2014a,israelachvili_11}
The higher the density, the smaller the available free volume for
the rigid cluster, thus the plateau becomes smaller and steeper. At
certain number density there is no more room to locate a particle
in the center of the cavity. The available space is so small that
a particle at the center would push the remaining ones out of bounds.
This is shown in the lower panel of Fig. \ref{fig:densN5}, which
exhibits an excluded volume region close to the center of the cavity.
This gives a clear insight of how, at high densities, the system conformation
and translation becomes dominated by the cavity shape. At high densities,
rigid clusters have low translational freedom and therefore their
constituent particles stand at approximately fixed distances from
the center. Cluster structure gets expressed on the density profiles
that show local maxima and minima. Increasing the temperature softens
the cluster, causing the local structure features to disappear, leaning
towards monotonous curves. Finally, at dissociation temperatures,
the depletion dominance gets clear and the density at the wall is
the highest in the profile. We note here a difficulty, intrinsically
related with the spherical shape of the cavity. The local properties
are hard to measure in the central region because it is poorly sampled.
Indeed, the sampling becomes poorer as $r$ decreases towards the
center of the pore. This effect is produced by the rapid reduction
of the sampled volume that produce large fluctuations in $\rho(r)$
and other quantities. These fluctuations can also be observed in Fig.
\ref{fig:denscomp1}.

\begin{figure}
\includegraphics[width=7cm]{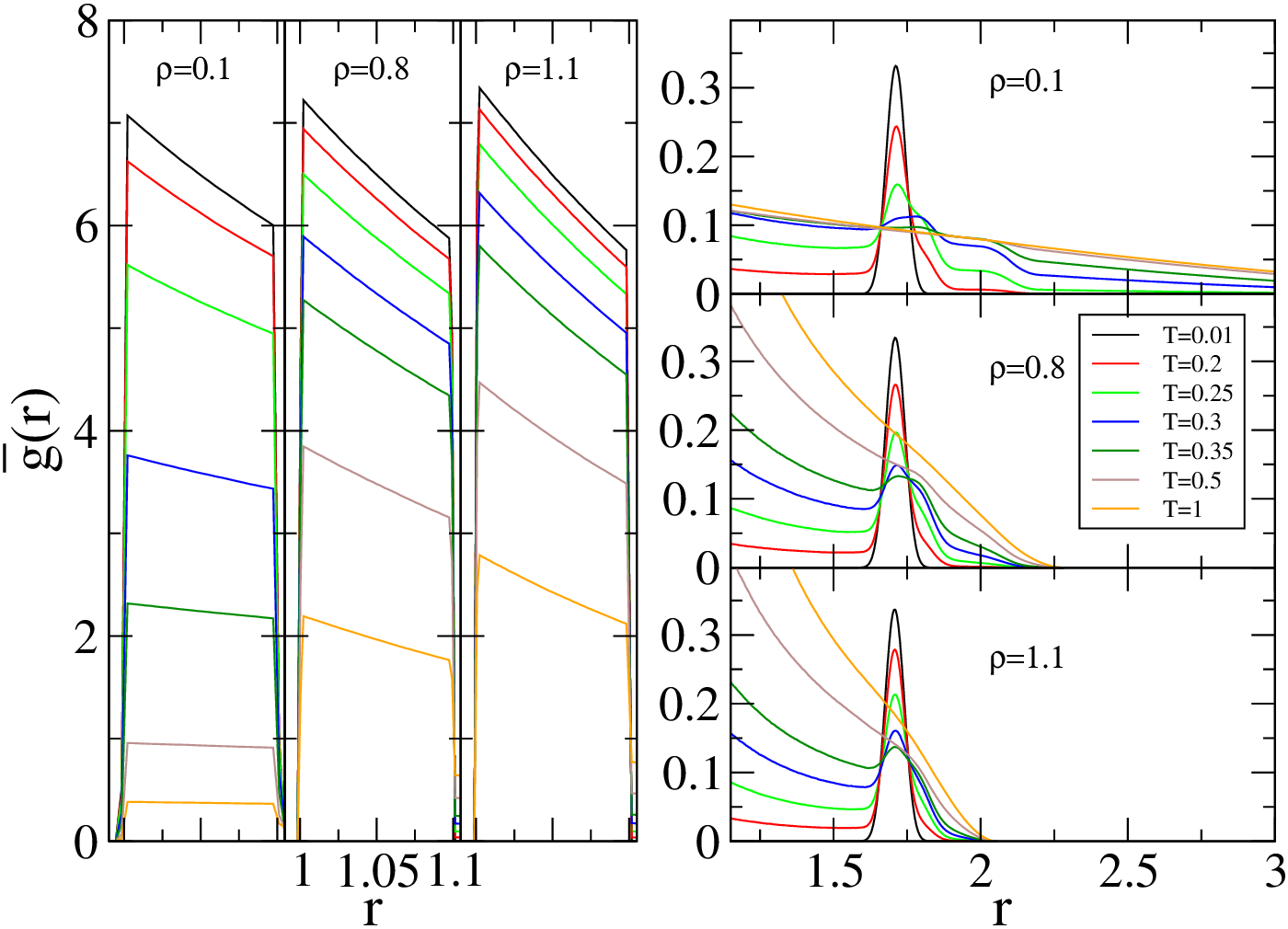}\protect\caption{\label{fig:pairN5.1}$\bar{g}(r)$ for $N=5$ at different temperatures
and representative number densities. Left panel: detail of the main
peak, right panel: detail of the second peak for $r\in\left[1.15;3\right]$.}
\end{figure}
In Figure \ref{fig:pairN5.1} a similar systematic approach is applied
for the $\bar{g}(r)$ curves under density variation. Each studied
density $\rho=0.1$, $0.8$ and $1.1$ corresponds to a cavity diameter
of $4.57$, $2.28$ and $2.05$, respectively. By increasing the temperature,
pair bonds are more likely to break and produce dissociations. The
odds of a separated pair to become together again is smaller at larger
free space. As a consequence, main peaks fall more abruptly at dissociation
temperatures and lower densities. At high densities, there is not
enough room for the particles to stay away from each other, which
forces an increase in the probability of finding pair separations
inside the well range, even if particles have a very high kinetic
energy, as compared to the well interaction energy $\varepsilon$.
At $\rho=1.1$ it is noticeable how cavity diameter is close to second
neighbor distance. For $N=5$ the rigid cluster geometry is also the
most compact the system can achieve, the cavity is only a slightly
larger than the smallest possible configuration. This implies that
the cluster as a whole is practically locked in the center, allowed
mostly only to rotate. As already observed in Figure \ref{fig:densN5},
at high densities and low temperatures, the particles of the rotating
rigid body maintain a stable distance from the center. For $\rho>1.1$,
the wall cavity squeezes the system, shortening the second neighbor
distance.

Despite the SW interaction is isoenergetic once inside the field range,
the main peak of $\bar{g}(r)$ shows a negative slope as can be observed
in Figs. \ref{fig:pairN2}, \ref{fig:paircomp1}, \ref{fig:paircomp2}
and \ref{fig:pairN5.1}. This implies that, within the interaction
range $r\in\left[1,1+\lambda\right]$, particles tend to be at closest
distance instead of near the external side of the well. Theoretical
results for $N=2$ (Eq. \ref{eq:pairN2}) points out that the slope
originates from the particle-cavity interaction. Basically the main
peak is an offset mounted on a decreasing function. This can be rationalized
by noting that a closer pair has more free space in the cavity than
a stretched one, increasing the traslational entropy. Also particle
collisions with a curved concave wall will, in average, tend to group
them together.

\subsection{Thermodynamic quantities\label{sub:ResThermodyn}}

We focus on four thermodynamic properties to characterize the system
as a whole. We analyze the pressure and the energy of the system,
that have robust definitions and are measured in a straightforward
way. Additionally, we study the surface tension and the surface adsorption,
which are intrinsically related with the inhomogeneous nature of the
system. These last quantities are more subtle and difficult to measure
by simulation.

\begin{figure}
\includegraphics[width=7cm]{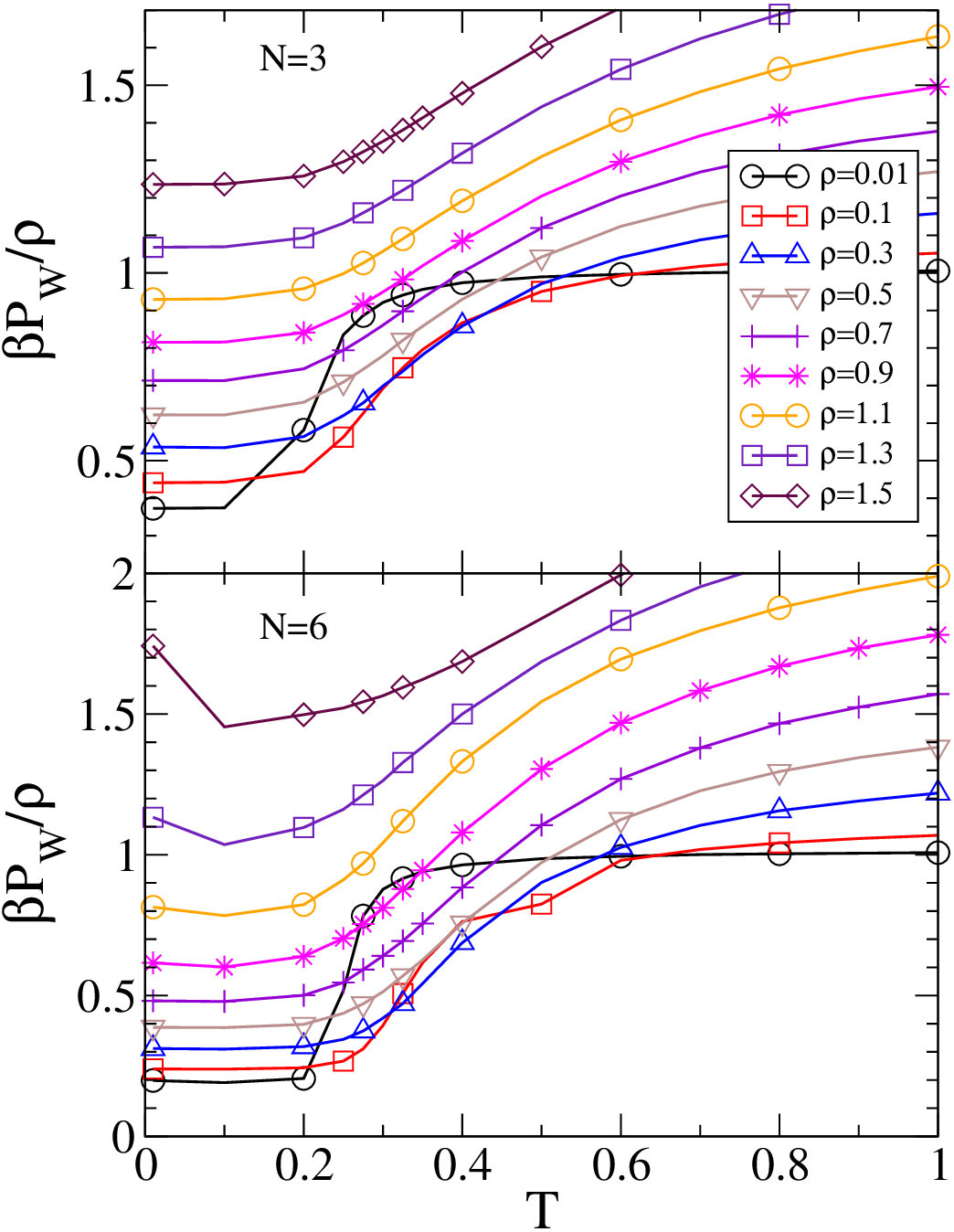}

\protect\caption{\label{fig:Pw1}Compressibility factor $\beta P_{w}/\rho$ as a function
of temperature, for different number densities for $N=3$ (Top panel)
and $N=6$ (Bottom panel).}
\end{figure}
Pressure on the wall is expected to increase with temperature and
number density, given that both parameters should increase the average
number of collisions on the wall. We work with the compresibility
factor $\beta P_{w}/\rho$ in Fig. \ref{fig:Pw1} to eliminate the
linear dependence, allowing to distinguish deviations from the ideal
case $\beta P_{w}/\rho=1$. Only the high temperature and very low
density cases follow the ideal behavior, when the relative well attraction
is too weak and excluded volume from the cores is negligible. Cluster
to dissociation temperatures are mediated by a sudden increase in
$\beta P_{w}/\rho$ and further saturation. The jump becomes smoother
for higher densities as a result of what has been pointed out from
$\bar{g}(r)$ analysis: with less space for separation, qualitative
differences between a cluster and unbounded particles are smaller.
The mentioned depletion emergence, for any density value at high $T$,
explains the asymptotic increase of $\beta P_{w}/\rho$ until the
HS limit.

At very low temperatures, the system is far away from the ideal gas
behavior because the relative potential well produces a cluster. In
this limit the system compressibility factor increases monotonically
for increasing density. We attribute this to the behavior of a unique
finite-size cluster allowed to stay in an effective volume $V_{\textrm{cluster}}<V$.
The single cluster behaves as an ideal gas, and thus, its compressibility
factor is $\beta PV_{\textrm{cluster}}=1$ i.e. $\beta P/\rho\simeq N^{-1}V/V_{\textrm{cluster}}$.
This explains the low-density and low-temperature values $\beta P_{w}/\rho\gtrsim0.33$
and $\beta P_{w}/\rho\gtrsim0.17$ for $N=3$ and $6$ respectively.
At intermediates temperatures, when the bond-breaking probability
is non-negligible, the case of low density raises much more rapidly
to its saturation value than those of higher densities. This could
be related to the strong reduction of recombination rate at lower
densities. Once a bond is broken the probability of the particles
to meet again is very small. This is not the case for intermediate
to high densities.

\begin{figure}
\includegraphics[width=7cm]{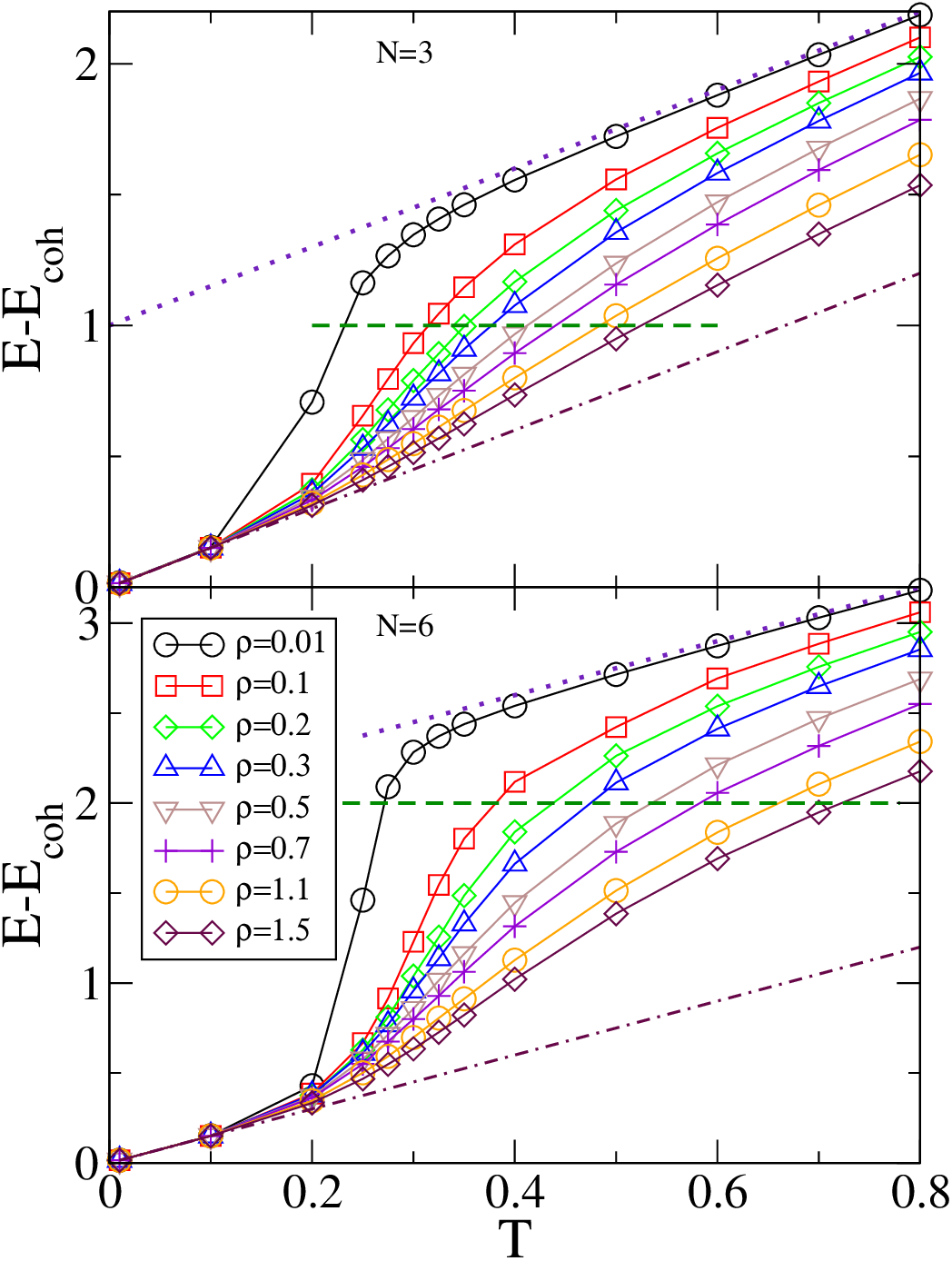}

\protect\caption{\label{fig:EtN3-6}$E-E_{\textrm{coh}}$ as a function of $T$ for
different number densities Top panel shows $N=3$ and Bottom panel
$N=6$ cases. The horizontal dashed line plots $\left|E_{\textrm{coh}}\right|=-E_{\textrm{coh}}$.}
\end{figure}
The mean energy per particle $E$ is the addition of the kinetic term
$3T/2$ driven by the temperature and the mean potential energy. For
low temperatures, where $\left|\phi_{N}\right|/N\gg3T/2$, the system
is in a rigid cluster state. Then, $E$ can be precisely calculated
as the number of bonded pairs for a given geometry. We call that value
cohesion energy $E_{\textrm{coh}}$, which is the lowest (fundamental)
possible energy of the system and we define it as the zero value in
Fig. \ref{fig:EtN3-6}, presenting the energy versus temperature.
Average energy per particle has a similar behavior to the one of the
pressure in Fig. \ref{fig:Pw1}. For low density, it presents an abrupt
increment in going from low to higher temperatures ($T\sim0.2$).
This jump agrees with the range of non-rigid cluster, ending at dissociation
temperatures ($T\sim0.3$). Then it follows a weak linear variation
for high temperatures, according to the equipartition theorem. It
is worth noting that for systems with a unique stable rigid cluster
geometry at low temperatures, all the curves must collapse to a single
with slope $3/2$, independently of the density. For reference, the
behavior of both, the pure kinetic energy and the shifted one (plus
$\left|E_{\textrm{coh}}\right|$), are also shown in dotted and dot-dashed
lines in Fig. \ref{fig:EtN3-6}. Rigid clusters constitute compact
structures, and increasing the density does not changes the number
of first neighbors. Higher density lines have lower values because
the particles are closer, so SW interactions are forced. The changes
produced with increasing temperature in each curve are more pronounced
for higher $N$, because the rigid cluster has more bounds (per particle)
to be broken. This feature is also shown for the different values
that takes $\left|E_{\textrm{coh}}\right|$, in dashed line. The line
also serves to visualize the condition $E=0$, which characterize
the equilibrium between potential and kinetic energies. This crossover
line separates two characteristic regions: one where the potential
energy dominates over the kinetic energy, proper of clusters, and
another one where kinetic energy dominates, a feature proper of gases.

\begin{figure}
\includegraphics[width=7cm]{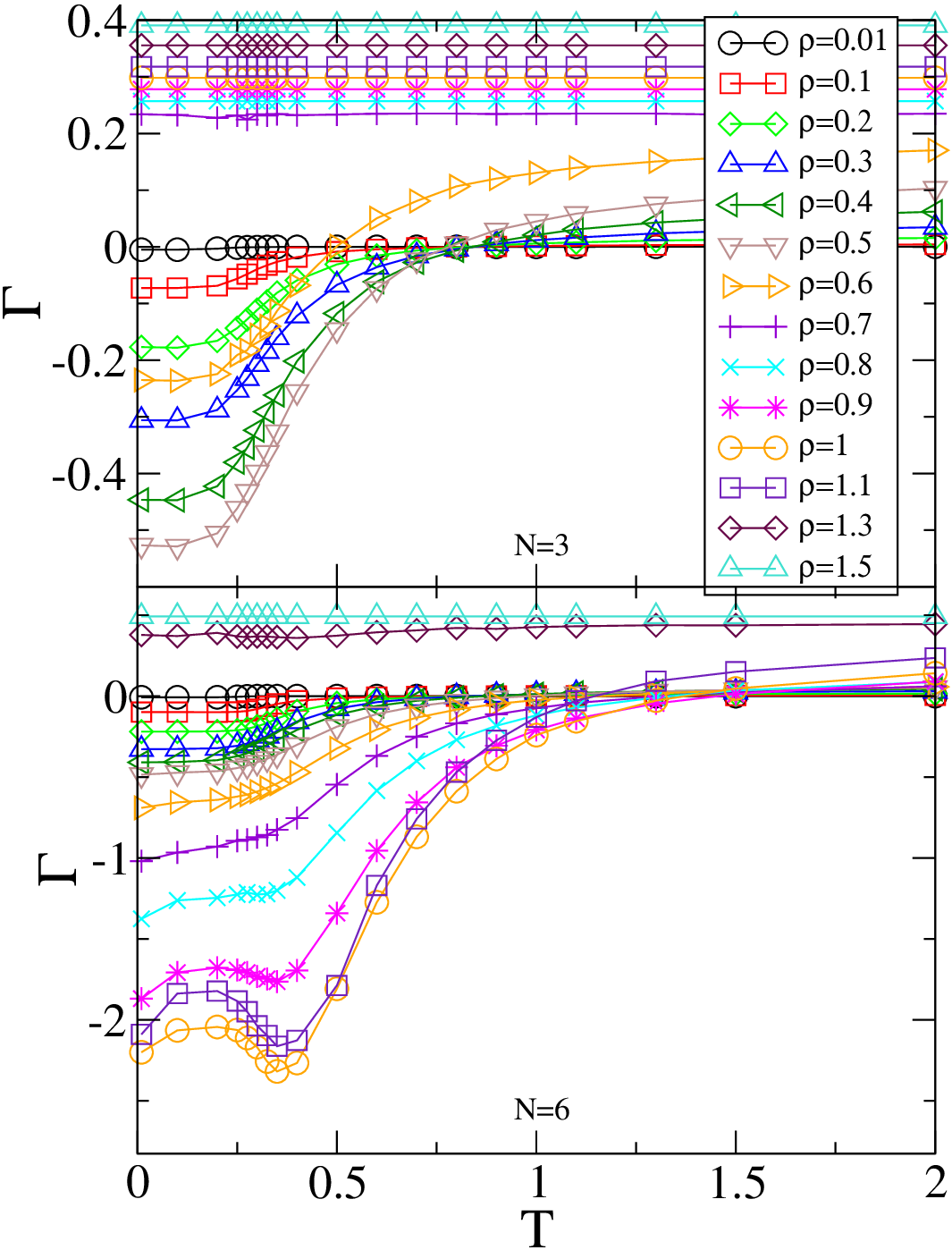}

\protect\caption{\label{fig:AdsorbN3-6}Adsorption $\Gamma$ as a function of temperature
for different number densities. The cases $N=3$ (Top panel) and $N=6$
(Bottom panel) are shown.}
\end{figure}

\begin{figure}
\includegraphics[width=7cm]{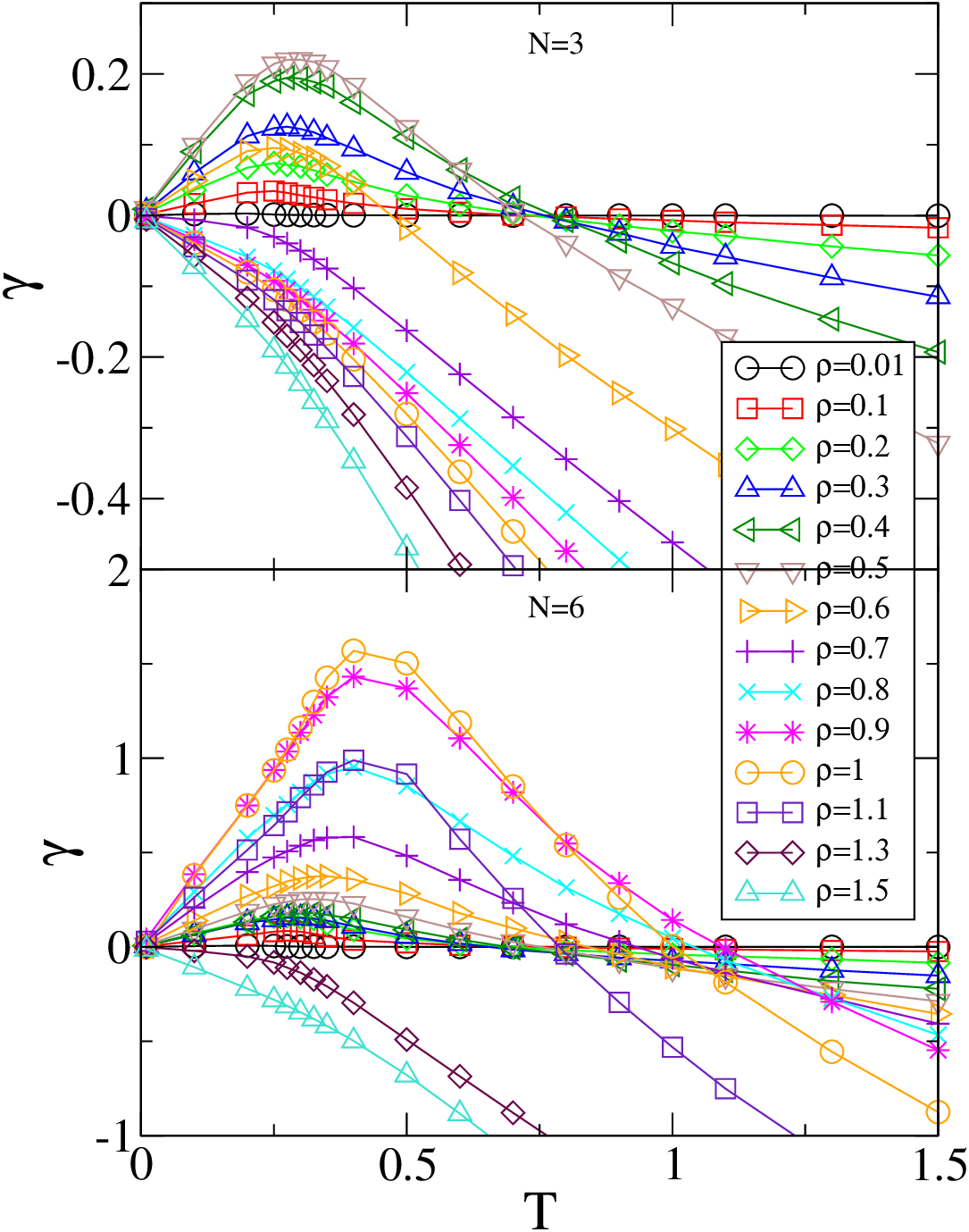}

\protect\caption{\label{fig:Tenss3-6}Surface tension as a function of $T$ for different
number densities. The cases $N=3$ (Top panel) and $N=6$ (Bottom
panel) are shown.}
\end{figure}
Surface adsorption $\Gamma$ and surface tension $\gamma$ are basic
properties used to characterize the inhomegeneity induced on the system
by the presence of walls. They are measured in the same way as a in
a previous work.\cite{Urrutia_2014a} We won't delve into details
and only give here the definition of $\Gamma$, and the expression
of $\gamma$, based on Laplace equation: $\Gamma=\left(\rho-\rho_{c}\right)\frac{V}{A}$
and $\gamma=\frac{P_{c}-P_{w}}{2}R_{\textrm{eff}}$. Here $\rho_{c}$
($P_{c}$) refers to the average density (pressure) near the center
of the cavity. These magnitudes are difficult to measure because one
must fix a criterion to choose the region where averages should be
done. The criterion must be applied for all the available range of
$\rho$ and $T$. Note that at large $R_{\textrm{eff}}$, the density
profiles attain a nearly constant value in the central region (\emph{plateau}).
At constant $N$, the density $\rho_{c}$ becomes higher for smaller
cavity radius until a limiting value, in which particles cannot freely
place themselves in the center. This central region progressively
turns into an excluded zone. It is important to note that at such
high packing values, the definition of $\rho_{c}$ is less accurate,
since there is not a clear distinction between the center region and
that in the vicinity of the wall. Adsorption is shown in Fig. \ref{fig:AdsorbN3-6}.
For low densities, adsorption is negative for small $T$ and then
experiences a jump around dissociation temperatures to flatten at
higher values of $T$. It becomes positive at high temperatures. For
high densities, adsorption is positive and nearly independent of the
temperature. The limiting cases of temperature are clearly identified.
For high temperature $\Gamma$ is positive and increase monotonously
with density. In the case of low temperature, for small densities
$\Gamma$ is negative and decreases with increasing $\rho$, up to
a certain minimum value. Further increment of $\rho$ produces a sudden
rise of $\Gamma$, that becomes positive. This behavior goes in line
with that observed in the density profiles, in Section \ref{sub:ResStructure}.
$\rho(r)$ presents an enhancement close to the wall for $\Gamma>0$
and an increment in the center for $\Gamma<0$.

Fig. \ref{fig:Tenss3-6} shows the surface tension $\gamma$ for three
(Top panel) and six particles (Bottom panel). At vanishing temperature
they start from 0, having then two distinctive behaviors. For low
to intermediate densities the $\gamma$ curves are positive at low
$T$. They start with positive slope, reach a maximum value, to become
negative at high temperatures. For high densities, $\gamma$ curves
are negative. They start with negative slope and decrease monotonously
with temperature. 

The two different characteristics of $\gamma$ can be rationalized
by considering that when clusters are favored at medium to low densities,
the system tends to get far from the cavity wall, having the cluster
size as its characteristic size. This minimizes the intrinsic area
of the system, going along with a negative adsorption and an increase
of density at the center of the cavity. At high enough temperatures,
the system behaves as HS particles, having the confining cavity as
a characteristic size, with an effective entropic attraction from
the wall, and a negative surface tension.\cite{Urrutia_2014a} The
curve that shows the global maximum of surface tension corresponds
to higher density for $N=6$ than for $N=3$. Also, the temperature
of those global maxima of $\gamma$ is shifted towards higher values.

The approximate linear dependence of $\gamma$ with $T$ for both
very low and high temperatures is explained by the expected hard sphere
limit where $\beta\gamma$ only depends on density. The behavior upon
variation of density is similar to that observed for $\Gamma$.

Some of the measured thermodynamic properties of the SW system can
be readily mapped to the AO model by the established relation between
$T$ and $\eta_{p}$. We expect one of these magnitudes to be the
pressure on the wall, once the energy scale is compensated {[}as in
the case of $\beta\phi_{\textrm{AO}}$ in Eq. (\ref{eq:potAO}){]},
thus $\beta P_{w}$ but also $\beta P_{w}/\rho$ that was plotted
in Fig. \ref{fig:Pw1} could be mapped. The measured surface tension
is similar to the pressure, it should be transformed to $\beta\gamma$.
A third magnitude is $\Gamma$ which does not scales with $T$ and
depends on characteristic features of $\rho(r)$. The case of energy
is more complicated because in the AO system the energy is purely
kinetic, and therefore the energy can not be mapped.

\subsection{Phase diagrams\label{sub:ResPhaseDiag}}

\begin{figure}
\begin{centering}
\includegraphics[width=0.98\columnwidth]{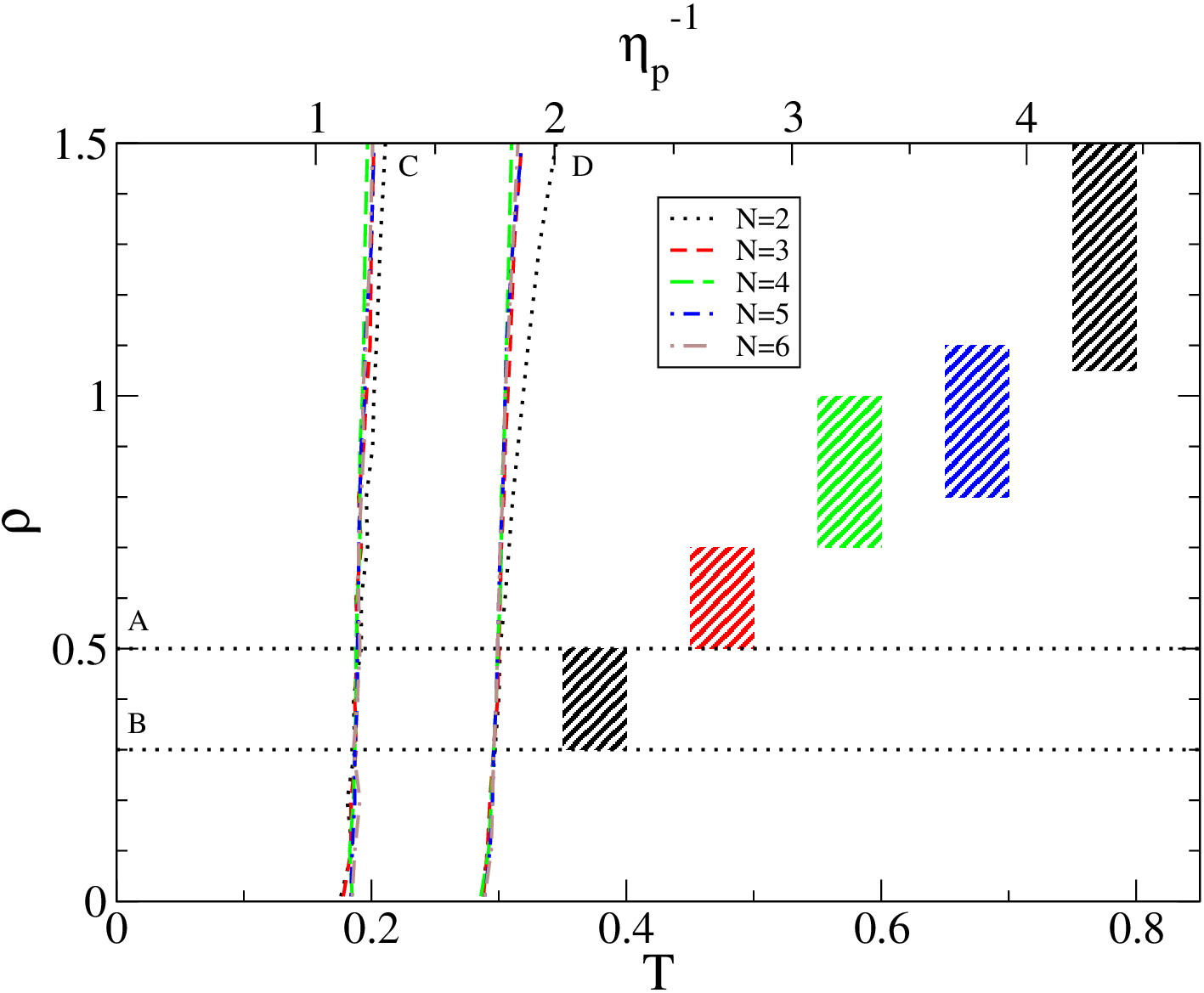}
\par\end{centering}

\protect\caption{Superimposed phase diagram for $N=2$ to $6$. Vertical lines come
from analyzing the pair distribution function. At the left side of
C, there is a hard cluster region. Between C and D we define a soft
or plastic cluster region and at the right side of C dissociation
starts to happen. Horizontal lines show density points where density
profiles are maximum (B), or vanish at the center (A). The rectangles
show the range of $\rho$ between A and B positions for every $N$.\label{fig:Superimposed-phase-diagram}}
\end{figure}
In Fig. \ref{fig:Superimposed-phase-diagram} we show the change of
the main characteristics of the phase diagram with the variation of
$N$. This summarizes variation of structural properties with $T$
and $\rho$. The vertical lines come from the analysis of the pair
distribution function, by comparing the height of the main peak $\bar{g}(r=1^{+})$
with that in the region next to the main peak $\bar{g}(r=1.1^{+})$,
and its connection with the qualitative behavior of the system, obtained
by direct visualization of the dynamics of the particles. C corresponds
to $\bar{g}(r=1.1^{+})/\bar{g}(r=1^{+})=0.005$ and D to $\bar{g}(r=1.1^{+})/\bar{g}(r=1^{+})=0.03$.
The relations between those points have been picked by noting that
those values match cluster softening (C) and dissociation process
(D). These lines divide temperature domains by particle conformation:
low temperatures until C mostly defined by particles gathering in
a hard cluster. Between C and D the system forms a soft cluster with
relative movements among particles. For higher temperatures, beyond
D, frequent dissociations are observed. The horizontal lines come
from the analysis of the main features of the density distribution.
Line A indicates the cases in which the density profile at the center
of the cavity $\rho(r=0)$ reaches its maximum, and line B when it
becomes zero {[}$\rho(r=0)=0${]}. The former case represents low
translational freedom for the particle that is at the center, while
for the later there is such a high density (or a small cavity), that
a particle can not be in the center due to the excluded volume. These
horizontal lines delimit the following regions: below B there is a
zone in which the system is moderately inhomogeneous. The region limited
by A and B corresponds to a strong reduction of freedom of motion
that reduces the occupation in the center. For densities beyond line
A the center of the cavity is an excluded volume (high confinement).
Note that a third horizontal line (not shown) fix the maximum density
of the confined system where it becomes completely caged. This density
can be calculated with a simple geometrical approach and varies with
$N$. In contrast to the vertical lines, the position of A and B are
strongly dependent on $N$.

\begin{figure}
\begin{centering}
\includegraphics[width=0.98\columnwidth]{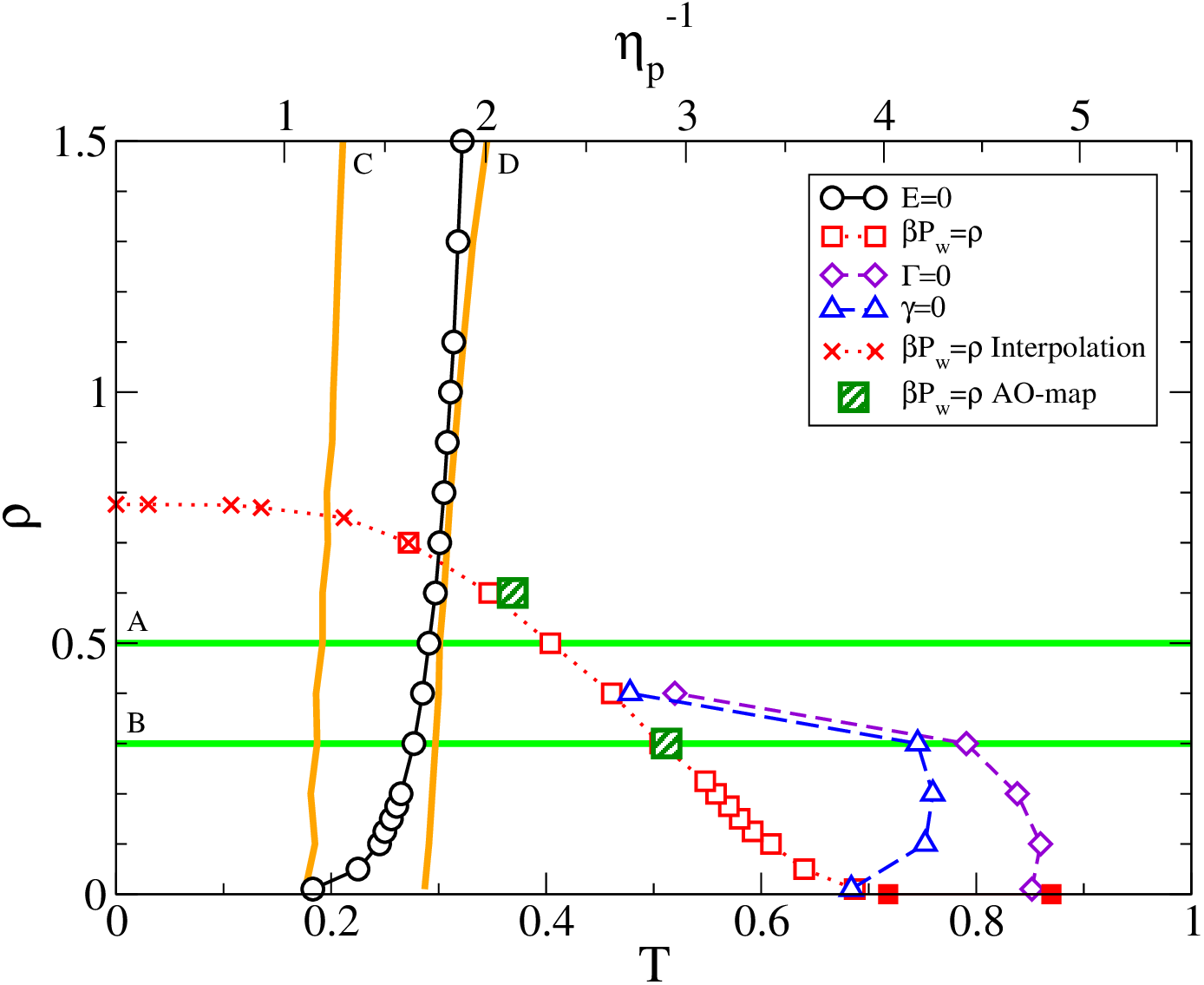}
\par\end{centering}

\protect\caption{\label{fig:Phase-diagramN2}Phase diagram for $N=2$ with characteristic
curves. A, B, C and D thick lines are the same from Fig.\ref{fig:Superimposed-phase-diagram}.
Additionally, curves indicating $E=0$ (black circles), $\beta P_{w}=\rho$
(open red squares); $\Gamma=0$ (green diamonds), and $\gamma=0$
(blue triangles) are shown. Red crosses present the values obtained
by interpolating the curves of $\beta P_{w}/\rho$ for specific density
values. These lie between points for which simulation results are
available thus allowing to complete the curve at the saturation area.
The two red filled squares show exact analytic values for $\beta P_{w}=\rho$
and $\Gamma=0$ at $\rho\rightarrow0$. Green bigger squares represent
the AO model results for $\beta P_{w}=\rho$. In the top x-axis the
scale of packing fraction of polymers in the AO model is presented.}
\end{figure}
\begin{figure}
\begin{centering}
\includegraphics[width=0.98\columnwidth]{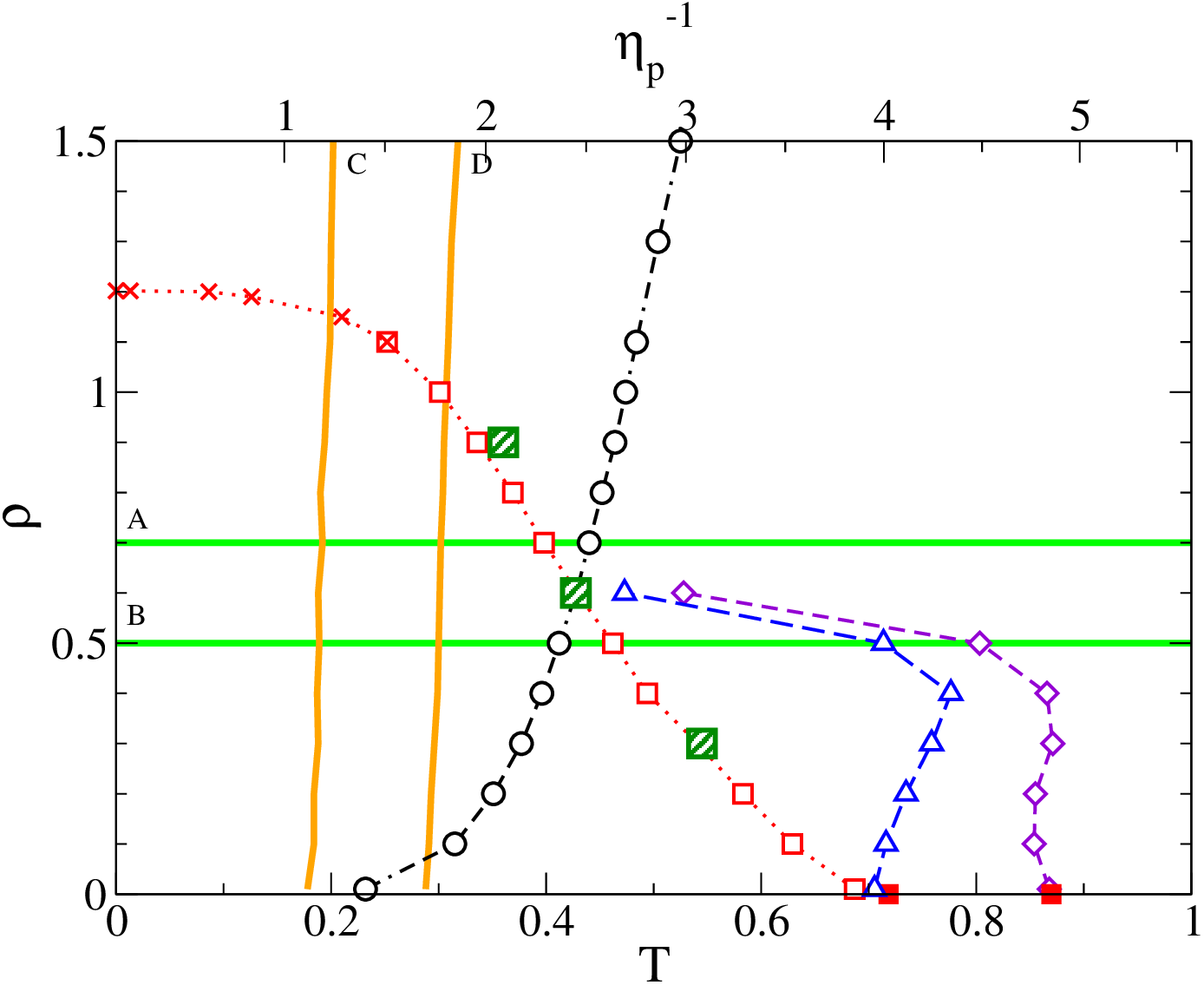}
\par\end{centering}

\protect\caption{\label{fig:Phase-diagramN3}Phase diagram for $N=3$ with characteristic
curves. Symbols, lines and top x-axis follow the same notation from
those of Fig.\ref{fig:Phase-diagramN2}.}
\end{figure}
\begin{figure}
\begin{centering}
\includegraphics[width=0.98\columnwidth]{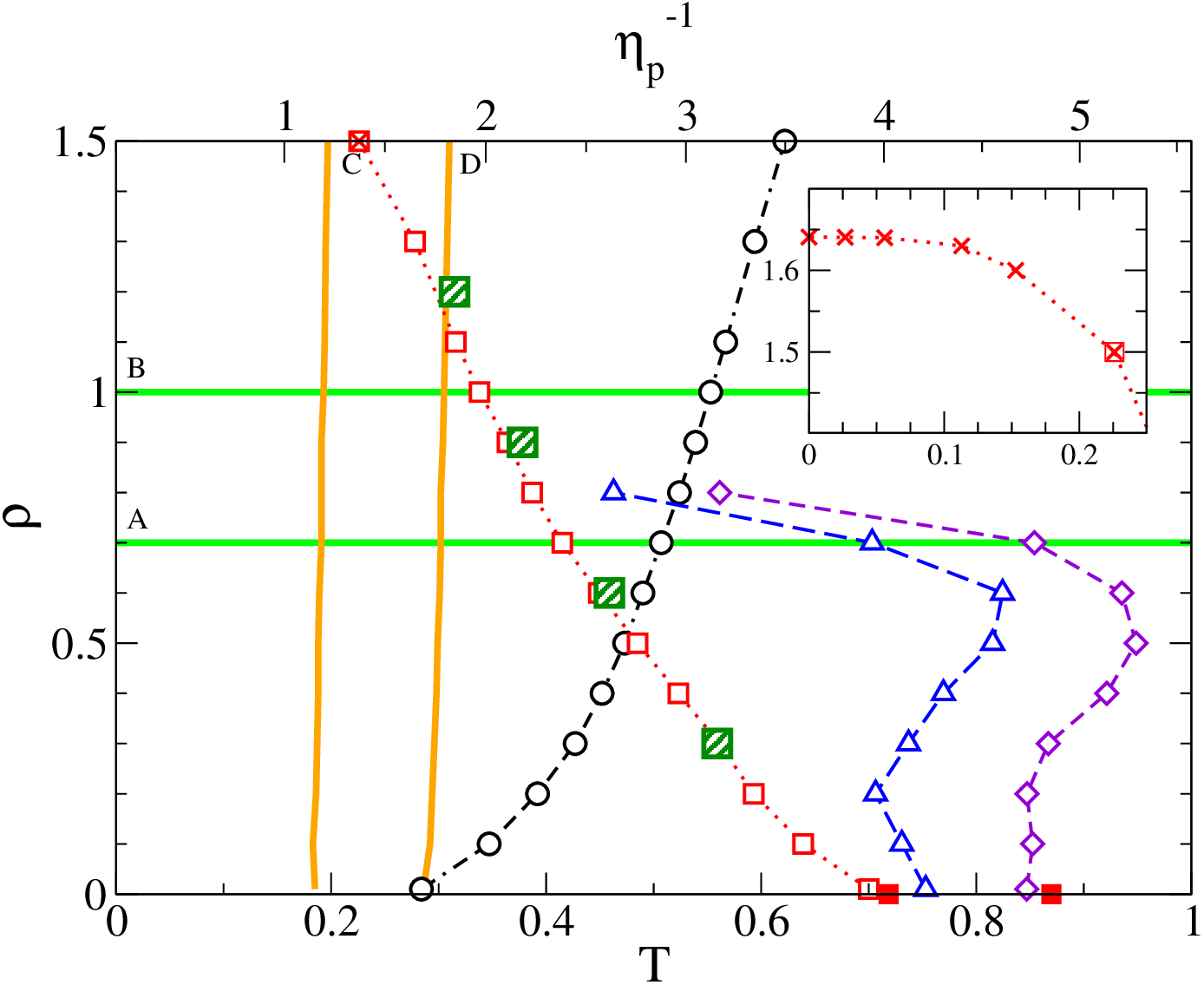}
\par\end{centering}

\protect\caption{\label{fig:Phase-diagramN4}Phase diagram for $N=4$ with characteristic
curves. Symbols, lines and top x-axis follow the same notation from
those of Fig.\ref{fig:Phase-diagramN2}. The inset shows details of
$\beta P_{w}=\rho$ at low temperatures.}
\end{figure}

\begin{figure}
\begin{centering}
\includegraphics[width=0.98\columnwidth]{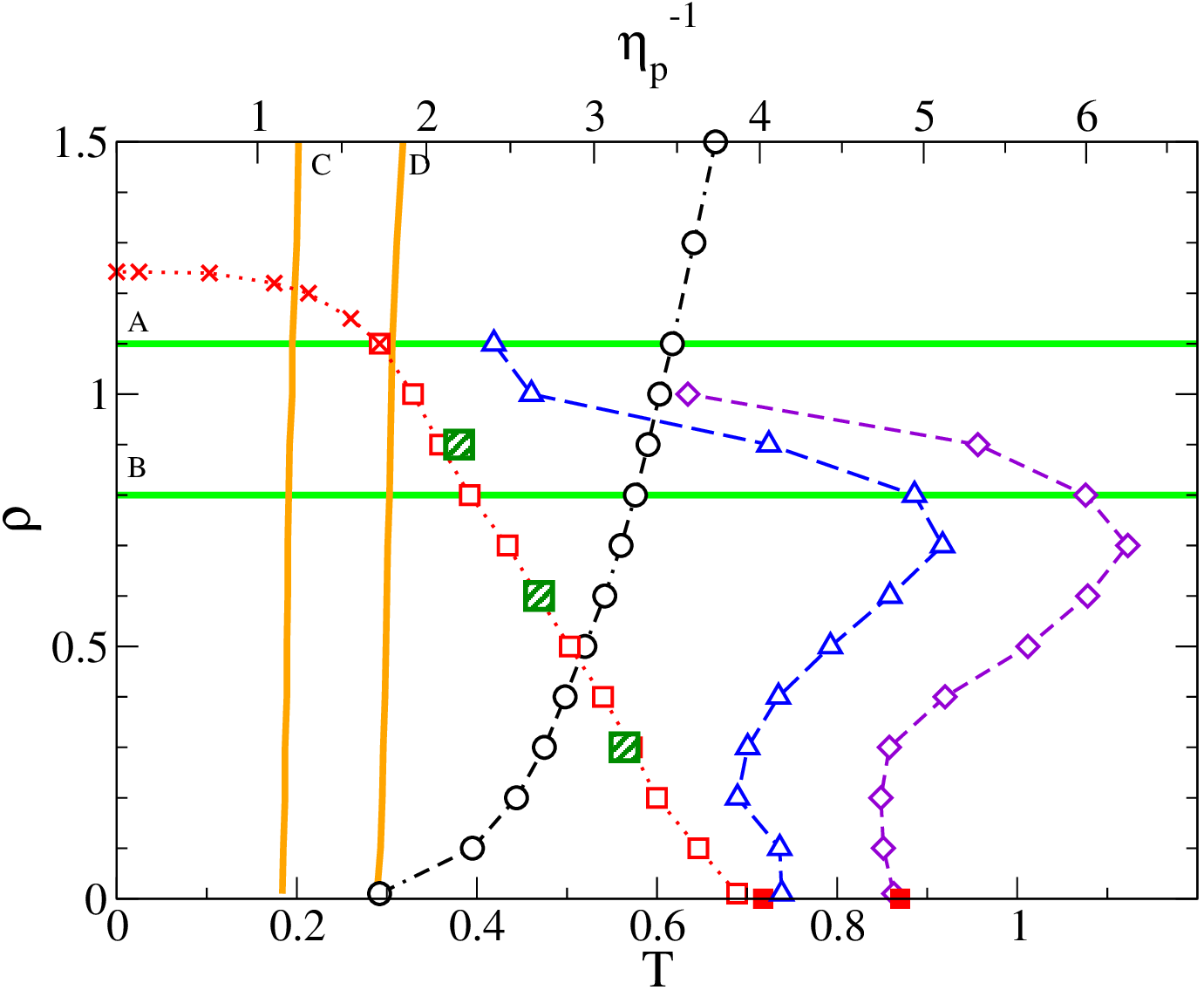}
\par\end{centering}

\protect\caption{\label{fig:Phase-diagramN5}Phase diagram for $N=5$ with characteristic
curves. Symbols, lines and top x-axis follow are the same notation
of those of Fig.\ref{fig:Phase-diagramN2}.}
\end{figure}

\begin{figure}
\begin{centering}
\includegraphics[width=0.98\columnwidth]{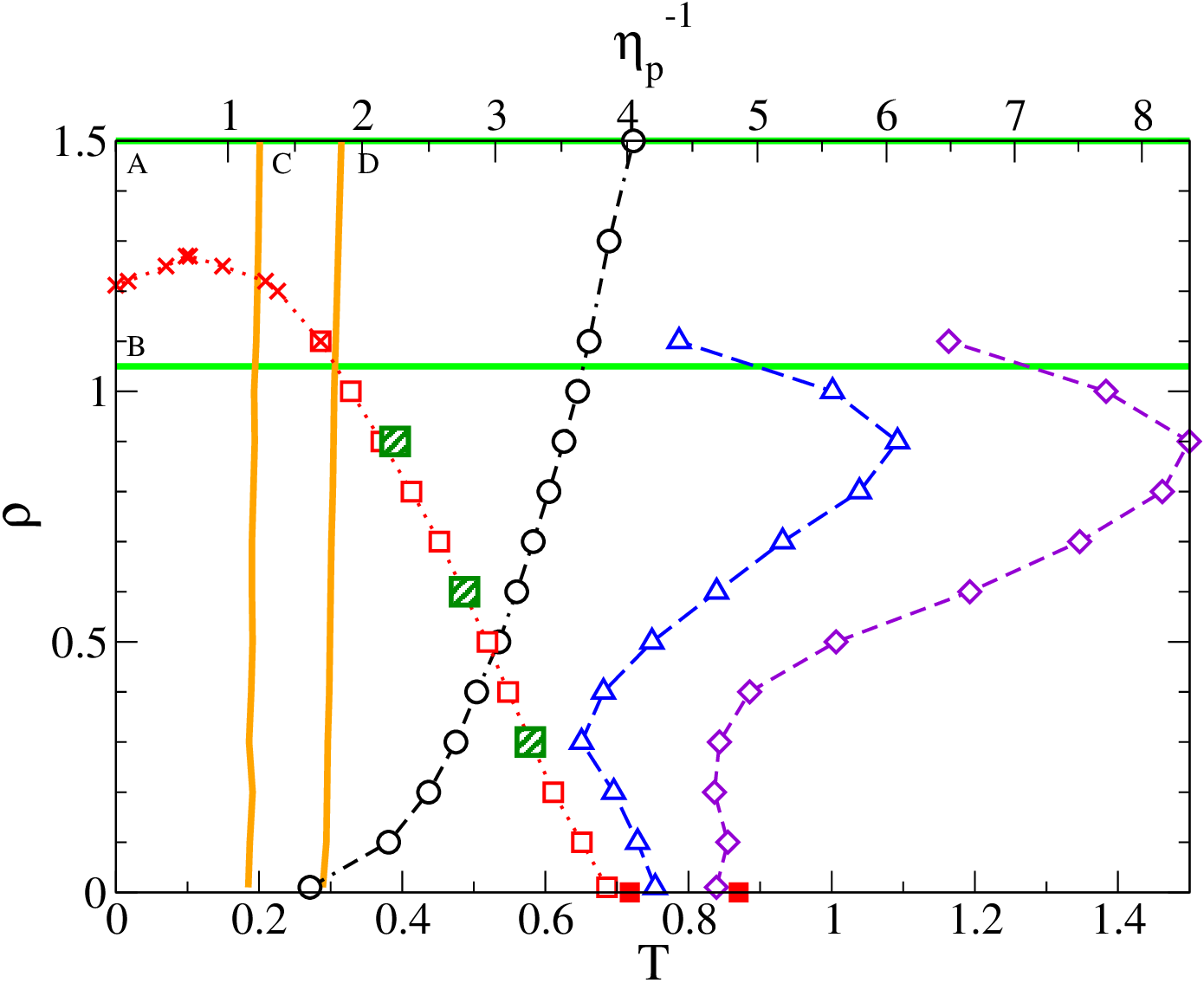}
\par\end{centering}

\protect\caption{\label{fig:Phase-diagramN6}Phase diagram for $N=6$ with characteristic
curves. Symbols, lines and top x-axis follow the notation of Fig.\ref{fig:Phase-diagramN2}.}
\end{figure}
In Figs. \ref{fig:Phase-diagramN2} to \ref{fig:Phase-diagramN6}
we present the phase diagram for the different number of particles
studied. It is shown a set of characteristic values of density and
temperature labeled with letter A, B, C and D. They separate regions
where the systems present different qualitative behavior. These regions
would represent different phases of the system in a macroscopic context.

Additionally, the scalars chosen to distinguish different qualitative
behavior of the system are the $(T,\rho)$ points on which: kinetic
energy is equal to potential energy (black circles), pressure on the
wall follows ideal gas equation of state (red open squares and crosses),
adsorption and surface tension change their signs (green diamonds
and blue triangles, respectively). We point out that the $(T,\rho)$
grid step is $0.1$, except for the cases of low temperatures (e.g.
$T=0.01$) and higher densities where the steps become more spaced.
In particular, low temperature $\beta P_{w}=\rho$ points (red crosses)
have been attained by interpolating $\beta P_{w}$ curves for specific
density values, whose variations are too small for our scale.

The $E=0$ curve (open circles) divides the phase space. Towards the
left, the (modulus of) potential energy is higher than the kinetic
energy and the opposite case is true, towards the right. In the region
with $E<0$ (at the left of $E=0$) the potential energy dominates.
This feature is characteristic of systems that spontaneously collapse
in cluster aggregates or condensed phases, but also of cold enough
compressed systems, where the available space is reduced. On the opposite,
when $E>0$ (at the right of $E=0$ curve) the kinetic energy dominates
and this is typical of systems of free particles, as the case of diluted
gases. Note that this last region includes a high enough temperature
and high density range, where particles stay together, as a consequence
of the strong confinement. The $E=0$ curve is specially sensitive
to changes in $N$, since it is affected by the average number of
bonds per particle. At low $N$, adding particles increases the maximum
number of bonds per particle, given that the system is too far away
from bulk case. As a result of this, the curve shifts towards higher
temperatures upon increasing $N$. This is specially observed comparing
Figures \ref{fig:Phase-diagramN2} and \ref{fig:Phase-diagramN3}.
The curve flattens for higher densities since smaller cavity forces
bond formation, leading to a fixed potential energy value of a tightly
packed cluster. This curve has not a counterpart for the colloid-polymer
mixture analyzed with the AO model. Even when there is an effective
potential between colloids, the system is athermal and the origin
of the potential is purely entropic. Both, colloids and polymers have
only kinetic energy, thus $E\neq0$ for non zero temperatures.

The curve $\beta P_{w}=\rho$ (open squares) indicates the points
of the $(T,\rho)$ where the EOS for the system pressure behaves like
the ideal gas pressure $P_{\textrm{id}}$. At the left side and below
of the $\beta P_{w}=\rho$ curve, $P_{w}<P_{\textrm{id}}$, the system
is undercompressed with respect to the ideal gas (at the same temperature
and number density). This region encloses the origin. Outside this
region $P_{w}>P_{\textrm{id}}$ the system is overcompressed in comparison
with the ideal gas. For low densities and high temperatures this is
expected as a consequence that depletion favors wall contact. Higher
densities force wall contact, independently of temperature. The theoretical
prediction based on the limit of large $V$ (low density) is shown
with a red square on $T$ axis. Notably, the obtained limiting temperature
fits the curve and is independent of $N$. From $N=3$ to $6$ all
these curves follow a similar behavior: from zero density up to $\rho\approx1$,
where $T\approx0.3$. The curves at lower temperatures show a strong
dependence with $N$. It is interesting to note that the critical
temperature of the vapor-liquid metastable transition in the studied
short-range SW system is $T=0.47$, which, for the largest values
of $N$, nearly coincides with the intersection of zero-energy and
ideal gas pressure curves.%
\footnote{These curves were extrapolated from surface tension data in Ref. \cite{LopezRendon_2006} %
}

The $\beta P_{w}=\rho$ curve was selected to verify the validity
of the corresponding-states mapping between SW and AO in confined
systems. Using Metropolis-Rosembluth Monte Carlo calculations\cite{Metropolis_1953,Rosenbluth_1954}
we have evaluated the density distribution of the AO system (for few
values of $\rho$), and used the contact theorem Eq. (\ref{eq:PwAO})
to evaluate $\beta P_{w}$. Thus, we seek for the value of $\eta_{p}$
that produce $\beta P_{w}=\rho$. Extended corresponding state law
{[}see Eq. (\ref{eq:AO-SW-map}){]} was used to evaluate the temperature
of the corresponding SW equivalent system. Calculated values are drawn
in green squares at Figs. \ref{fig:Phase-diagramN2} to \ref{fig:Phase-diagramN6}.
The obtained values of $\eta_{p}^{-1}$ are given at the top horizontal
x axis, while the equivalent temperature of the SW system can be read
at the bottom axis. We found a general coincidence between Monte Carlo
results for AO and simulation results for SW, along all the analyzed
range. At lower temperatures, near clusterization, the mapping between
both systems becomes poorer. This is expected, because the use of
the extended law of corresponding states is not documented for freezing
temperatures.

Zero surface adsorption and surface tension curves indicate where
there is no excess in surface of particle concentration or free energy,
respectively. As it was mentioned in Sec. \ref{sub:ResThermodyn},
both are difficult to measure. $\Gamma$ and $\gamma$ reveal a strong
reduction of accuracy at very low density where the system is quasi-homogeneous
and both quantities become very small. This makes specially hard to
measure the value of $T$ at which $\varGamma=0$ and $\gamma=0$
for $\rho\rightarrow0$. In addition, given that $\Gamma$ and $\gamma$
become ill defined at high densities we do not evaluate zero surface
excess in this case. In summary, the results presented here give a
general idea of the position and shape of both curves in the $(T,\rho)$
plane. Adsorption illustrates the relation between the density of
particles at the center of the cavity and those closer to the wall.
Towards the left side, where $\Gamma<0$, the particles are more likely
to be at the center of the cavity. For $\Gamma>0$, at the right side,
particles favor positions close to the wall.

\section{Conclusions\label{sec:End}}

In this work we studied thoroughly the properties of few colloidal
particles confined in a spherical cavity. We adopt the short-range
square well model and provide a deep characterization of the structural
and thermal properties of systems of 2, 3, 4, 5 and 6 particles in
a spherical pore for the complete relevant ranges of density and temperature.
Additionally, we compare the simulations with exact results for the
case $N=2$. Applying an extension of the corresponding states law
to confined systems, we establish a mapping between the square well
system and the AO model for effective interactions between colloids
in a polymer colloid mixture. We also developed the statistical mechanical
approach to systems of few particles, SW and AO, in confinement and
map the pressure on the wall at a given temperature in the SW system
to the equivalent packing fraction of polymers in the AO system.

The structure of the system ranging from low density to almost caging
of the particles in the cavity was characterized through the pair
correlation function and density profiles for the entire relevant
range of temperatures. Thermal bulk properties such as energy and
pressure on the wall were calculated and characterized. Different
effects of confinement were also studied, identifying their energetic
or entropic origin and focusing on the inhomogeneities present in
the system. Surface properties were analyzed with quantities reminiscent
of surface tension and adsorption in macroscopic counterparts of the
square well system.

We characterized the morphology of these systems, defining different
regions of similar behavior and criteria to provide phase diagrams
in the $(T,\rho)$ plane, for the different number of particles. In
this phase diagram, we identified temperature regions where the system
behaves as a rigid cluster, as a plastic cluster, and a region where
the system dissociates, up to the limit of hard-sphere-like behavior
at very high temperatures. In the density domain, we recognized regions
with different degrees of inhomogeneity which can be classified in
the following categories: low-to-moderate, moderate to excluded volume,
and excluded-volume to caging regions. We defined several characteristic
curves in the phase diagram, such as that of zero energy, ideal gas
pressure, zero adsorption and zero surface tension. These lines delimit
meaningful references in the phase diagram, that were used as a complement
for the analysis.
\begin{acknowledgments}
Financial support through grants PICT-2011-1887, PICT-2011-1217, PIP
112-200801-00403, INN-CNEA 2011, PICT-E 2014, is gratefully acknowledged.
\end{acknowledgments}

\appendix

\section{Effective potential and AO partition function\label{sec:Appendix}}

We show here how to transform Eq. (\ref{eq:SGCPF}) in Eq. (\ref{eq:SGCPF2}).
To this end, we analyze the term $\sum_{N_{p}}\frac{z_{p}^{N_{p}}}{N_{p}!}Z_{N,N_{p}}$
in Eq. (\ref{eq:SGCPF}), where the (canonical ensemble) colloid-polymer
mixture CI reads
\begin{equation}
Z_{N,N_{p}}=\int_{V^{N}}\int_{V_{p}^{Np}}e^{-\beta\left(\phi_{cc}+\phi_{cp}\right)}d\mathbf{r}_{p}^{N_{p}}d\mathbf{r}_{c}^{N}\:,\label{aeq:ZNNp}
\end{equation}
with $\phi_{cc}=\sum_{i,j}\phi_{ij}^{(cc)}$, $\phi_{cp}=\sum_{i}^{(c)}\sum_{k}^{(p)}\phi_{ik}^{(cp)}$.
$\phi_{ij}^{(cc)}$, $\phi_{ik}^{(cp)}$ are the spherically symmetric
pair potentials. Note that the region $\mathcal{C}$ where the center
of colloids are confined (with volume $V$) is different to the region
$\mathcal{P}$ where the center of polymers lies (with volume $V_{p}$).
In fact, the boundary of $\mathcal{P}$ must be placed in a region
where the polymers reach their bulk properties. For the AO system
with $q<0.1547$ confined in a spherical pore, the smallest region
$\mathcal{P}$ is an sphere with radius $R_{0}+\sigma_{p}/2$.

Polymers behave as ideal gas particles. If we fix the position of
the colloids, they exert a fixed external potential to the polymers
and thus
\begin{equation}
\int_{V_{p}^{Np}}e^{-\beta\phi_{cp}}d\mathbf{r}_{p}^{N_{p}}=\biggl(\int_{V_{p}}e^{-\beta\phi_{cp}}d\mathbf{r}_{p}\biggr)^{N_{p}}=Z_{\otimes}^{N_{p}}\:,\label{aeq:Zx}
\end{equation}
where $Z_{\otimes}$ is the CI of one polymer in $V_{p}$ at fixed
colloids. Furthermore, 
\begin{eqnarray}
\sum_{N_{p}}\frac{z_{p}^{N_{p}}}{N_{p}!}Z_{N,N_{p}} & = & \int_{V^{N}}e^{-\beta\phi_{cc}}\sum_{N_{p}}\frac{z_{p}^{N_{p}}}{N_{p}!}Z_{\otimes}^{N_{p}}d\mathbf{r}_{c}^{N}\:,\label{aeq:sumR1}\\
 & = & \int_{V^{N}}e^{\bigl(-\beta\phi_{cc}+z_{p}Z_{\otimes}\bigr)}d\mathbf{r}_{c}^{N}\:,\label{aeq:sumR2}
\end{eqnarray}
and thus we can simply analyze the case of one polymer. We introduce
the Mayer function for the colloid/polymer Boltzmann statistical weight
$e_{i}=\exp\bigl(-\beta\phi_{i1}^{(cp)}\bigr)=1+f_{i}$ in $Z_{\otimes}$
to obtain
\begin{equation}
Z_{\otimes}=\int_{V_{p}}\Bigl(1+\sum_{i}f_{i}+\sum_{<ij>}f_{i}f_{j}+\ldots\Bigr)d\mathbf{r}_{p}\:,\label{aeq:Zx1}
\end{equation}
where higher order terms are products of three or more functions $f$
concerning the position of three or more colloids. In this integrand
$f_{i}$ is minus one for $\mathbf{r}_{p}$ such that the (center-to-center)
$i$th-colloid to polymer distance fulfills $x_{i}<\sigma+\sigma_{p}$
and otherwise is zero. $f_{i}f_{j}$ is one if $\mathbf{r}_{p}$ fulfills
both $x_{i}<\sigma+\sigma_{p}$ and $x_{j}<\sigma+\sigma_{p}$ and
is zero otherwise, and so on. Once integrated, $Z_{\otimes}$ takes
the form
\begin{equation}
V_{p}-\frac{4\pi}{3}\sigma^{3}\left(1+q\right)^{3}N+\sum_{<ij>}V_{o}\bigl(r_{ij}\bigr)+\ldots\label{aeq:Zx2}
\end{equation}
Here $V_{o}\bigl(r_{ij}\bigr)$ is the overlap volume between two
spheres with radius $\bigl(\sigma+\sigma_{p}\bigr)/2$ and extra terms
include the overlap of at least three spheres. Turning to the integrand
of Eq. (\ref{aeq:sumR2}), we utilize the identities $z_{p}=\rho_{p}$
and $z_{p}V_{p}=N_{p}$ to obtain
\begin{equation}
\exp\left(N_{p}-N_{p}^{x}\right)\exp\Bigl[-\beta\sum_{<ij>}\phi_{\textrm{HS}}\bigl(r_{ij}\bigr)+\rho_{p}\sum_{<ij>}V_{o}\bigl(r_{ij}\bigr)\Bigr]\:,\label{aeq:end}
\end{equation}
with $N_{p}^{x}=\rho_{p}v_{\textrm{exc}}N$ and $v_{\textrm{exc}}=\frac{4\pi}{3}\sigma^{3}\left(1+q\right)^{3}$.
In addition, we neglected higher order terms in Eq. (\ref{aeq:end}).
These terms are null if $q<0.1547$. For $q\apprge0.1547$, including
the case $q=0.2$ analyzed in the present work, one expects that three-body
contribution will be negligible in comparison with two-body terms.
Naturally, $\Xi_{p,h}=\exp N_{p}$ and $\beta\phi_{\textrm{AO}}(r)=\beta\phi_{\textrm{HS}}(r)-\rho_{p}V_{o}(r)$,
where $\beta\phi_{\textrm{AO}}(r)$ is the same expression given in
Eq. (\ref{eq:potAO}). Therefore, Eq. (\ref{aeq:sumR2}) is $Z_{N}^{(\textrm{AO})}$.
This demonstrates the equivalence between Eq. (\ref{eq:SGCPF}) and
Eq. (\ref{eq:SGCPF2}).

The described procedure can be generalized in several ways. It is
not restricted to the spherical pore, and thus, it applies to other
pore geometries like cylinders, cuboids, slits, single walls, etc.
Further extensions include the case of non-free polymers where both,
colloids and polymers, are confined, the AO model with $q\apprge0.1547$,
and also other non-AO systems with more general interaction potentials.
It can also be readily applied to systems of AO particles in spaces
with dimensions other than three (discs and hyper-spheres), with $V_{o}\bigl(r\bigr)$
taken from Ref.\cite{Urrutia_2010}.


\begin{thebibliography}{60}%
\makeatletter
\providecommand \@ifxundefined [1]{%
 \@ifx{#1\undefined}
}%
\providecommand \@ifnum [1]{%
 \ifnum #1\expandafter \@firstoftwo
 \else \expandafter \@secondoftwo
 \fi
}%
\providecommand \@ifx [1]{%
 \ifx #1\expandafter \@firstoftwo
 \else \expandafter \@secondoftwo
 \fi
}%
\providecommand \natexlab [1]{#1}%
\providecommand \enquote  [1]{``#1''}%
\providecommand \bibnamefont  [1]{#1}%
\providecommand \bibfnamefont [1]{#1}%
\providecommand \citenamefont [1]{#1}%
\providecommand \href@noop [0]{\@secondoftwo}%
\providecommand \href [0]{\begingroup \@sanitize@url \@href}%
\providecommand \@href[1]{\@@startlink{#1}\@@href}%
\providecommand \@@href[1]{\endgroup#1\@@endlink}%
\providecommand \@sanitize@url [0]{\catcode `\\12\catcode `\$12\catcode
  `\&12\catcode `\#12\catcode `\^12\catcode `\_12\catcode `\%12\relax}%
\providecommand \@@startlink[1]{}%
\providecommand \@@endlink[0]{}%
\providecommand \url  [0]{\begingroup\@sanitize@url \@url }%
\providecommand \@url [1]{\endgroup\@href {#1}{\urlprefix }}%
\providecommand \urlprefix  [0]{URL }%
\providecommand \Eprint [0]{\href }%
\providecommand \doibase [0]{http://dx.doi.org/}%
\providecommand \selectlanguage [0]{\@gobble}%
\providecommand \bibinfo  [0]{\@secondoftwo}%
\providecommand \bibfield  [0]{\@secondoftwo}%
\providecommand \translation [1]{[#1]}%
\providecommand \BibitemOpen [0]{}%
\providecommand \bibitemStop [0]{}%
\providecommand \bibitemNoStop [0]{.\EOS\space}%
\providecommand \EOS [0]{\spacefactor3000\relax}%
\providecommand \BibitemShut  [1]{\csname bibitem#1\endcsname}%
\let\auto@bib@innerbib\@empty
\bibitem [{\citenamefont {Wang}\ \emph {et~al.}(2011)\citenamefont {Wang},
  \citenamefont {Chai}, \citenamefont {Wang}, \citenamefont {Li}, \citenamefont
  {Liu}, \citenamefont {Zhang}, \citenamefont {Su},\ and\ \citenamefont
  {Liao}}]{Wang_11}%
  \BibitemOpen
  \bibfield  {author} {\bibinfo {author} {\bibfnamefont {T.-T.}\ \bibnamefont
  {Wang}}, \bibinfo {author} {\bibfnamefont {F.}~\bibnamefont {Chai}}, \bibinfo
  {author} {\bibfnamefont {C.-G.}\ \bibnamefont {Wang}}, \bibinfo {author}
  {\bibfnamefont {L.}~\bibnamefont {Li}}, \bibinfo {author} {\bibfnamefont
  {H.-Y.}\ \bibnamefont {Liu}}, \bibinfo {author} {\bibfnamefont {L.-Y.}\
  \bibnamefont {Zhang}}, \bibinfo {author} {\bibfnamefont {Z.-M.}\ \bibnamefont
  {Su}}, \ and\ \bibinfo {author} {\bibfnamefont {Y.}~\bibnamefont {Liao}},\
  }\href {\doibase http://dx.doi.org/10.1016/j.jcis.2011.02.023} {\bibfield
  {journal} {\bibinfo  {journal} {Journal of Colloid and Interface Science}\
  }\textbf {\bibinfo {volume} {358}},\ \bibinfo {pages} {109 } (\bibinfo {year}
  {2011})}\BibitemShut {NoStop}%
\bibitem [{\citenamefont {Weeks}(2012)}]{Weeks_2012}%
  \BibitemOpen
  \bibfield  {author} {\bibinfo {author} {\bibfnamefont {E.~R.}\ \bibnamefont
  {Weeks}},\ }\href {\doibase 10.1126/science.1228952} {\bibfield  {journal}
  {\bibinfo  {journal} {Science}\ }\textbf {\bibinfo {volume} {338}},\ \bibinfo
  {pages} {55} (\bibinfo {year} {2012})}\BibitemShut {NoStop}%
\bibitem [{\citenamefont {Sacanna}\ \emph {et~al.}(2010)\citenamefont
  {Sacanna}, \citenamefont {Irvine}, \citenamefont {Chaikin},\ and\
  \citenamefont {Pine}}]{Sacanna_2010}%
  \BibitemOpen
  \bibfield  {author} {\bibinfo {author} {\bibfnamefont {S.}~\bibnamefont
  {Sacanna}}, \bibinfo {author} {\bibfnamefont {W.~T.~M.}\ \bibnamefont
  {Irvine}}, \bibinfo {author} {\bibfnamefont {P.~M.}\ \bibnamefont {Chaikin}},
  \ and\ \bibinfo {author} {\bibfnamefont {D.~J.}\ \bibnamefont {Pine}},\
  }\href {\doibase 10.1038/nature08906} {\bibfield  {journal} {\bibinfo
  {journal} {Nature}\ }\textbf {\bibinfo {volume} {464}},\ \bibinfo {pages}
  {575} (\bibinfo {year} {2010})}\BibitemShut {NoStop}%
\bibitem [{\citenamefont {Meng}\ \emph {et~al.}(2010)\citenamefont {Meng},
  \citenamefont {Arkus}, \citenamefont {Brenner},\ and\ \citenamefont
  {Manoharan}}]{Meng_2010}%
  \BibitemOpen
  \bibfield  {author} {\bibinfo {author} {\bibfnamefont {G.}~\bibnamefont
  {Meng}}, \bibinfo {author} {\bibfnamefont {N.}~\bibnamefont {Arkus}},
  \bibinfo {author} {\bibfnamefont {M.~P.}\ \bibnamefont {Brenner}}, \ and\
  \bibinfo {author} {\bibfnamefont {V.~N.}\ \bibnamefont {Manoharan}},\ }\href
  {\doibase 10.1126/science.1181263} {\bibfield  {journal} {\bibinfo  {journal}
  {Science}\ }\textbf {\bibinfo {volume} {327}},\ \bibinfo {pages} {560}
  (\bibinfo {year} {2010})}\BibitemShut {NoStop}%
\bibitem [{\citenamefont {Lu}\ \emph {et~al.}(2008)\citenamefont {Lu},
  \citenamefont {Zaccarelli}, \citenamefont {Ciulla}, \citenamefont
  {Schofield}, \citenamefont {Sciortino},\ and\ \citenamefont
  {Weitz}}]{Lu_2008}%
  \BibitemOpen
  \bibfield  {author} {\bibinfo {author} {\bibfnamefont {P.~J.}\ \bibnamefont
  {Lu}}, \bibinfo {author} {\bibfnamefont {E.}~\bibnamefont {Zaccarelli}},
  \bibinfo {author} {\bibfnamefont {F.}~\bibnamefont {Ciulla}}, \bibinfo
  {author} {\bibfnamefont {A.~B.}\ \bibnamefont {Schofield}}, \bibinfo {author}
  {\bibfnamefont {F.}~\bibnamefont {Sciortino}}, \ and\ \bibinfo {author}
  {\bibfnamefont {D.~A.}\ \bibnamefont {Weitz}},\ }\href {\doibase
  10.1038/nature06931} {\bibfield  {journal} {\bibinfo  {journal} {Nature}\
  }\textbf {\bibinfo {volume} {453}},\ \bibinfo {pages} {499} (\bibinfo {year}
  {2008})}\BibitemShut {NoStop}%
\bibitem [{\citenamefont {Wang}\ \emph {et~al.}(2012)\citenamefont {Wang},
  \citenamefont {Wang}, \citenamefont {Peng}, \citenamefont {Zheng},\ and\
  \citenamefont {Han}}]{Wang_2012}%
  \BibitemOpen
  \bibfield  {author} {\bibinfo {author} {\bibfnamefont {Z.}~\bibnamefont
  {Wang}}, \bibinfo {author} {\bibfnamefont {F.}~\bibnamefont {Wang}}, \bibinfo
  {author} {\bibfnamefont {Y.}~\bibnamefont {Peng}}, \bibinfo {author}
  {\bibfnamefont {Z.}~\bibnamefont {Zheng}}, \ and\ \bibinfo {author}
  {\bibfnamefont {Y.}~\bibnamefont {Han}},\ }\href {\doibase
  10.1126/science.1224763} {\bibfield  {journal} {\bibinfo  {journal}
  {Science}\ }\textbf {\bibinfo {volume} {338}},\ \bibinfo {pages} {87}
  (\bibinfo {year} {2012})}\BibitemShut {NoStop}%
\bibitem [{\citenamefont {Aarts}, \citenamefont {Schmidt},\ and\ \citenamefont
  {Lekkerkerker}(2004)}]{Aarts_2004_b}%
  \BibitemOpen
  \bibfield  {author} {\bibinfo {author} {\bibfnamefont {D.~G. A.~L.}\
  \bibnamefont {Aarts}}, \bibinfo {author} {\bibfnamefont {M.}~\bibnamefont
  {Schmidt}}, \ and\ \bibinfo {author} {\bibfnamefont {H.~N.~W.}\ \bibnamefont
  {Lekkerkerker}},\ }\href {\doibase 10.1126/science.1097116} {\bibfield
  {journal} {\bibinfo  {journal} {Science}\ }\textbf {\bibinfo {volume}
  {304}},\ \bibinfo {pages} {847} (\bibinfo {year} {2004})}\BibitemShut
  {NoStop}%
\bibitem [{\citenamefont {Anderson}\ and\ \citenamefont
  {Lekkerkerker}(2002)}]{Anderson_2002}%
  \BibitemOpen
  \bibfield  {author} {\bibinfo {author} {\bibfnamefont {V.~J.}\ \bibnamefont
  {Anderson}}\ and\ \bibinfo {author} {\bibfnamefont {H.~N.~W.}\ \bibnamefont
  {Lekkerkerker}},\ }\href {\doibase 10.1038/416811a} {\bibfield  {journal}
  {\bibinfo  {journal} {Nature}\ }\textbf {\bibinfo {volume} {416}},\ \bibinfo
  {pages} {811} (\bibinfo {year} {2002})}\BibitemShut {NoStop}%
\bibitem [{\citenamefont {de~Hoog}\ \emph {et~al.}(2001)\citenamefont
  {de~Hoog}, \citenamefont {Kegel}, \citenamefont {van Blaaderen},\ and\
  \citenamefont {Lekkerkerker}}]{deHoog_2001}%
  \BibitemOpen
  \bibfield  {author} {\bibinfo {author} {\bibfnamefont {E.~H.~A.}\
  \bibnamefont {de~Hoog}}, \bibinfo {author} {\bibfnamefont {W.~K.}\
  \bibnamefont {Kegel}}, \bibinfo {author} {\bibfnamefont {A.}~\bibnamefont
  {van Blaaderen}}, \ and\ \bibinfo {author} {\bibfnamefont {H.~N.~W.}\
  \bibnamefont {Lekkerkerker}},\ }\href {\doibase 10.1103/PhysRevE.64.021407}
  {\bibfield  {journal} {\bibinfo  {journal} {Phys. Rev. E}\ }\textbf {\bibinfo
  {volume} {64}},\ \bibinfo {pages} {021407} (\bibinfo {year}
  {2001})}\BibitemShut {NoStop}%
\bibitem [{\citenamefont {Royall}, \citenamefont {Poon},\ and\ \citenamefont
  {Weeks}(2013)}]{Royall_2013}%
  \BibitemOpen
  \bibfield  {author} {\bibinfo {author} {\bibfnamefont {C.~P.}\ \bibnamefont
  {Royall}}, \bibinfo {author} {\bibfnamefont {W.~C.~K.}\ \bibnamefont {Poon}},
  \ and\ \bibinfo {author} {\bibfnamefont {E.~R.}\ \bibnamefont {Weeks}},\
  }\href {\doibase 10.1039/C2SM26245B} {\bibfield  {journal} {\bibinfo
  {journal} {Soft Matter}\ }\textbf {\bibinfo {volume} {9}},\ \bibinfo {pages}
  {17} (\bibinfo {year} {2013})}\BibitemShut {NoStop}%
\bibitem [{\citenamefont {Royall}, \citenamefont {Louis},\ and\ \citenamefont
  {Tanaka}(2007)}]{Royall_2007}%
  \BibitemOpen
  \bibfield  {author} {\bibinfo {author} {\bibfnamefont {C.~P.}\ \bibnamefont
  {Royall}}, \bibinfo {author} {\bibfnamefont {A.~A.}\ \bibnamefont {Louis}}, \
  and\ \bibinfo {author} {\bibfnamefont {H.}~\bibnamefont {Tanaka}},\ }\href
  {\doibase 10.1063/1.2755962} {\bibfield  {journal} {\bibinfo  {journal} {The
  Journal of Chemical Physics}\ }\textbf {\bibinfo {volume} {127}},\ \bibinfo
  {eid} {044507} (\bibinfo {year} {2007})}\BibitemShut {NoStop}%
\bibitem [{\citenamefont {Urrutia}\ and\ \citenamefont
  {Pastorino}(2014)}]{Urrutia_2014a}%
  \BibitemOpen
  \bibfield  {author} {\bibinfo {author} {\bibfnamefont {I.}~\bibnamefont
  {Urrutia}}\ and\ \bibinfo {author} {\bibfnamefont {C.}~\bibnamefont
  {Pastorino}},\ }\href {\doibase http://dx.doi.org/10.1063/1.4896221}
  {\bibfield  {journal} {\bibinfo  {journal} {The Journal of Chemical Physics}\
  }\textbf {\bibinfo {volume} {141}},\ \bibinfo {eid} {124905} (\bibinfo {year}
  {2014})}\BibitemShut {NoStop}%
\bibitem [{\citenamefont {Urrutia}\ and\ \citenamefont
  {Castelletti}(2012)}]{Urrutia_2012}%
  \BibitemOpen
  \bibfield  {author} {\bibinfo {author} {\bibfnamefont {I.}~\bibnamefont
  {Urrutia}}\ and\ \bibinfo {author} {\bibfnamefont {G.}~\bibnamefont
  {Castelletti}},\ }\href {\doibase 10.1063/1.4729249} {\bibfield  {journal}
  {\bibinfo  {journal} {The Journal of Chemical Physics}\ }\textbf {\bibinfo
  {volume} {136}},\ \bibinfo {eid} {224509} (\bibinfo {year}
  {2012})}\BibitemShut {NoStop}%
\bibitem [{\citenamefont {Urrutia}(2011)}]{Urrutia_2011_b}%
  \BibitemOpen
  \bibfield  {author} {\bibinfo {author} {\bibfnamefont {I.}~\bibnamefont
  {Urrutia}},\ }\href {\doibase 10.1063/1.3609796} {\bibfield  {journal}
  {\bibinfo  {journal} {The Journal of Chemical Physics}\ }\textbf {\bibinfo
  {volume} {135}},\ \bibinfo {pages} {024511} (\bibinfo {year} {2011})},\
  \bibinfo {note} {erratum: ibid. 135(9), 099903 (2011)}\BibitemShut {NoStop}%
\bibitem [{\citenamefont {Acedo}\ and\ \citenamefont
  {Santos}(2001)}]{Acedo_2001}%
  \BibitemOpen
  \bibfield  {author} {\bibinfo {author} {\bibfnamefont {L.}~\bibnamefont
  {Acedo}}\ and\ \bibinfo {author} {\bibfnamefont {A.}~\bibnamefont {Santos}},\
  }\href {\doibase 10.1063/1.1384419} {\bibfield  {journal} {\bibinfo
  {journal} {The Journal of Chemical Physics}\ }\textbf {\bibinfo {volume}
  {115}},\ \bibinfo {pages} {2805} (\bibinfo {year} {2001})}\BibitemShut
  {NoStop}%
\bibitem [{\citenamefont {Li}\ \emph {et~al.}(2014)\citenamefont {Li},
  \citenamefont {Sun}, \citenamefont {Chen}, \citenamefont {Wang},\ and\
  \citenamefont {Cai}}]{Li_2014}%
  \BibitemOpen
  \bibfield  {author} {\bibinfo {author} {\bibfnamefont {L.}~\bibnamefont
  {Li}}, \bibinfo {author} {\bibfnamefont {F.}~\bibnamefont {Sun}}, \bibinfo
  {author} {\bibfnamefont {Z.}~\bibnamefont {Chen}}, \bibinfo {author}
  {\bibfnamefont {L.}~\bibnamefont {Wang}}, \ and\ \bibinfo {author}
  {\bibfnamefont {J.}~\bibnamefont {Cai}},\ }\href {\doibase 10.1063/1.4891799}
  {\bibfield  {journal} {\bibinfo  {journal} {The Journal of Chemical Physics}\
  }\textbf {\bibinfo {volume} {141}},\ \bibinfo {eid} {054905} (\bibinfo {year}
  {2014})}\BibitemShut {NoStop}%
\bibitem [{\citenamefont {Rivera-Torres}\ \emph {et~al.}(2013)\citenamefont
  {Rivera-Torres}, \citenamefont {del R\'{\i}o}, \citenamefont
  {Esp\'{\i}ndola-Heredia}, \citenamefont {Kolafa},\ and\ \citenamefont
  {Malijevsk\'{y}}}]{RiveraTorres_2013}%
  \BibitemOpen
  \bibfield  {author} {\bibinfo {author} {\bibfnamefont {S.}~\bibnamefont
  {Rivera-Torres}}, \bibinfo {author} {\bibfnamefont {F.}~\bibnamefont {del
  R\'{\i}o}}, \bibinfo {author} {\bibfnamefont {R.}~\bibnamefont
  {Esp\'{\i}ndola-Heredia}}, \bibinfo {author} {\bibfnamefont {J.}~\bibnamefont
  {Kolafa}}, \ and\ \bibinfo {author} {\bibfnamefont {A.}~\bibnamefont
  {Malijevsk\'{y}}},\ }\href {\doibase 10.1016/j.molliq.2012.12.005} {\bibfield
   {journal} {\bibinfo  {journal} {Journal of Molecular Liquids}\ }\textbf
  {\bibinfo {volume} {185}},\ \bibinfo {pages} {44} (\bibinfo {year}
  {2013})}\BibitemShut {NoStop}%
\bibitem [{\citenamefont {Esp\'{\i}ndola-Heredia}, \citenamefont {R\'{\i}o},\
  and\ \citenamefont {Malijevsk\'{y}}(2009)}]{EspindolaHeredia_2009}%
  \BibitemOpen
  \bibfield  {author} {\bibinfo {author} {\bibfnamefont {R.}~\bibnamefont
  {Esp\'{\i}ndola-Heredia}}, \bibinfo {author} {\bibfnamefont {F.~d.}\
  \bibnamefont {R\'{\i}o}}, \ and\ \bibinfo {author} {\bibfnamefont
  {A.}~\bibnamefont {Malijevsk\'{y}}},\ }\href {\doibase 10.1063/1.3054361}
  {\bibfield  {journal} {\bibinfo  {journal} {The Journal of Chemical Physics}\
  }\textbf {\bibinfo {volume} {130}},\ \bibinfo {eid} {024509} (\bibinfo {year}
  {2009})}\BibitemShut {NoStop}%
\bibitem [{\citenamefont {Vortler}, \citenamefont {Schafer},\ and\
  \citenamefont {Smith}(2008)}]{Vortler_2008}%
  \BibitemOpen
  \bibfield  {author} {\bibinfo {author} {\bibfnamefont {H.~L.}\ \bibnamefont
  {Vortler}}, \bibinfo {author} {\bibfnamefont {K.}~\bibnamefont {Schafer}}, \
  and\ \bibinfo {author} {\bibfnamefont {W.~R.}\ \bibnamefont {Smith}},\ }\href
  {\doibase 10.1021/jp073726r} {\bibfield  {journal} {\bibinfo  {journal} {The
  Journal of Physical Chemistry B}\ }\textbf {\bibinfo {volume} {112}},\
  \bibinfo {pages} {4656} (\bibinfo {year} {2008})}\BibitemShut {NoStop}%
\bibitem [{\citenamefont {L\'{o}pez-Rend\'{o}n}, \citenamefont {Reyes},\ and\
  \citenamefont {Orea}(2006)}]{LopezRendon_2006}%
  \BibitemOpen
  \bibfield  {author} {\bibinfo {author} {\bibfnamefont {R.}~\bibnamefont
  {L\'{o}pez-Rend\'{o}n}}, \bibinfo {author} {\bibfnamefont {Y.}~\bibnamefont
  {Reyes}}, \ and\ \bibinfo {author} {\bibfnamefont {P.}~\bibnamefont {Orea}},\
  }\href {\doibase 10.1063/1.2338307} {\bibfield  {journal} {\bibinfo
  {journal} {The Journal of Chemical Physics}\ }\textbf {\bibinfo {volume}
  {125}},\ \bibinfo {eid} {084508} (\bibinfo {year} {2006})}\BibitemShut
  {NoStop}%
\bibitem [{\citenamefont {Kiselev}, \citenamefont {Ely},\ and\ \citenamefont
  {Elliott}(2006)}]{Kiselev_2006}%
  \BibitemOpen
  \bibfield  {author} {\bibinfo {author} {\bibfnamefont {S.~B.}\ \bibnamefont
  {Kiselev}}, \bibinfo {author} {\bibfnamefont {J.~F.}\ \bibnamefont {Ely}}, \
  and\ \bibinfo {author} {\bibfnamefont {J.~R.}\ \bibnamefont {Elliott}},\
  }\href {\doibase 10.1080/00268970600808340} {\bibfield  {journal} {\bibinfo
  {journal} {Molecular Physics}\ }\textbf {\bibinfo {volume} {104}},\ \bibinfo
  {pages} {2545} (\bibinfo {year} {2006})}\BibitemShut {NoStop}%
\bibitem [{\citenamefont {Liu}, \citenamefont {Garde},\ and\ \citenamefont
  {Kumar}(2005)}]{Liu_2005}%
  \BibitemOpen
  \bibfield  {author} {\bibinfo {author} {\bibfnamefont {H.}~\bibnamefont
  {Liu}}, \bibinfo {author} {\bibfnamefont {S.}~\bibnamefont {Garde}}, \ and\
  \bibinfo {author} {\bibfnamefont {S.}~\bibnamefont {Kumar}},\ }\href
  {\doibase 10.1063/1.2085051} {\bibfield  {journal} {\bibinfo  {journal} {The
  Journal of Chemical Physics}\ }\textbf {\bibinfo {volume} {123}},\ \bibinfo
  {eid} {174505} (\bibinfo {year} {2005})}\BibitemShut {NoStop}%
\bibitem [{\citenamefont {Khanpour}(2011)}]{Mehrdad_2011}%
  \BibitemOpen
  \bibfield  {author} {\bibinfo {author} {\bibfnamefont {M.}~\bibnamefont
  {Khanpour}},\ }\href {\doibase 10.1103/PhysRevE.83.021203} {\bibfield
  {journal} {\bibinfo  {journal} {Phys. Rev. E}\ }\textbf {\bibinfo {volume}
  {83}},\ \bibinfo {pages} {021203} (\bibinfo {year} {2011})}\BibitemShut
  {NoStop}%
\bibitem [{\citenamefont {Hartskeerl}(2009)}]{Hartskeerl2009}%
  \BibitemOpen
  \bibfield  {author} {\bibinfo {author} {\bibfnamefont {T.}~\bibnamefont
  {Hartskeerl}},\ }\emph {\bibinfo {title} {Crystallization and Glassy
  Behaviour in Short-range Attractive Square-well Fluids}},\ \href
  {http://igitur-archive.library.uu.nl/student-theses/2011-0628-200551/UUindex.html}
  {Ph.D. thesis} (\bibinfo {year} {2009})\BibitemShut {NoStop}%
\bibitem [{\citenamefont {Neitsch}\ and\ \citenamefont
  {Klapp}(2013)}]{Neitsch_2013}%
  \BibitemOpen
  \bibfield  {author} {\bibinfo {author} {\bibfnamefont {H.}~\bibnamefont
  {Neitsch}}\ and\ \bibinfo {author} {\bibfnamefont {S.~H.~L.}\ \bibnamefont
  {Klapp}},\ }\href {\doibase 10.1063/1.4790406} {\bibfield  {journal}
  {\bibinfo  {journal} {The Journal of Chemical Physics}\ }\textbf {\bibinfo
  {volume} {138}},\ \bibinfo {eid} {064904} (\bibinfo {year}
  {2013})}\BibitemShut {NoStop}%
\bibitem [{\citenamefont {Armas-P\'{e}rez}, \citenamefont {Quintana-H},\ and\
  \citenamefont {Chapela}(2013)}]{ArmasPerez_2013}%
  \BibitemOpen
  \bibfield  {author} {\bibinfo {author} {\bibfnamefont {J.~C.}\ \bibnamefont
  {Armas-P\'{e}rez}}, \bibinfo {author} {\bibfnamefont {J.}~\bibnamefont
  {Quintana-H}}, \ and\ \bibinfo {author} {\bibfnamefont {G.~A.}\ \bibnamefont
  {Chapela}},\ }\href {\doibase 10.1063/1.4775342} {\bibfield  {journal}
  {\bibinfo  {journal} {The Journal of Chemical Physics}\ }\textbf {\bibinfo
  {volume} {138}},\ \bibinfo {eid} {044508} (\bibinfo {year}
  {2013})}\BibitemShut {NoStop}%
\bibitem [{\citenamefont {Huang}\ \emph {et~al.}(2010)\citenamefont {Huang},
  \citenamefont {Chen}, \citenamefont {Singh},\ and\ \citenamefont
  {Kwak}}]{Huang_2010}%
  \BibitemOpen
  \bibfield  {author} {\bibinfo {author} {\bibfnamefont {H.~C.}\ \bibnamefont
  {Huang}}, \bibinfo {author} {\bibfnamefont {W.~W.}\ \bibnamefont {Chen}},
  \bibinfo {author} {\bibfnamefont {J.~K.}\ \bibnamefont {Singh}}, \ and\
  \bibinfo {author} {\bibfnamefont {S.~K.}\ \bibnamefont {Kwak}},\ }\href
  {\doibase 10.1063/1.3429741} {\bibfield  {journal} {\bibinfo  {journal} {The
  Journal of Chemical Physics}\ }\textbf {\bibinfo {volume} {132}},\ \bibinfo
  {eid} {224504} (\bibinfo {year} {2010})}\BibitemShut {NoStop}%
\bibitem [{\citenamefont {Jana}, \citenamefont {Singh},\ and\ \citenamefont
  {Kwak}(2009)}]{Jana_2009}%
  \BibitemOpen
  \bibfield  {author} {\bibinfo {author} {\bibfnamefont {S.}~\bibnamefont
  {Jana}}, \bibinfo {author} {\bibfnamefont {J.~K.}\ \bibnamefont {Singh}}, \
  and\ \bibinfo {author} {\bibfnamefont {S.~K.}\ \bibnamefont {Kwak}},\ }\href
  {\doibase 10.1063/1.3148884} {\bibfield  {journal} {\bibinfo  {journal} {The
  Journal of Chemical Physics}\ }\textbf {\bibinfo {volume} {130}},\ \bibinfo
  {eid} {214707} (\bibinfo {year} {2009})}\BibitemShut {NoStop}%
\bibitem [{\citenamefont {Zhang}\ and\ \citenamefont
  {Wang}(2006)}]{Zhang_2006}%
  \BibitemOpen
  \bibfield  {author} {\bibinfo {author} {\bibfnamefont {X.}~\bibnamefont
  {Zhang}}\ and\ \bibinfo {author} {\bibfnamefont {W.}~\bibnamefont {Wang}},\
  }\href {\doibase 10.1103/PhysRevE.74.062601} {\bibfield  {journal} {\bibinfo
  {journal} {Phys. Rev. E}\ }\textbf {\bibinfo {volume} {74}},\ \bibinfo
  {pages} {062601} (\bibinfo {year} {2006})}\BibitemShut {NoStop}%
\bibitem [{\citenamefont {Pagan}\ and\ \citenamefont
  {Gunton}(2005)}]{Pagan_2005}%
  \BibitemOpen
  \bibfield  {author} {\bibinfo {author} {\bibfnamefont {D.~L.}\ \bibnamefont
  {Pagan}}\ and\ \bibinfo {author} {\bibfnamefont {J.~D.}\ \bibnamefont
  {Gunton}},\ }\href {\doibase 10.1063/1.1890925} {\bibfield  {journal}
  {\bibinfo  {journal} {The Journal of Chemical Physics}\ }\textbf {\bibinfo
  {volume} {122}},\ \bibinfo {eid} {184515} (\bibinfo {year}
  {2005})}\BibitemShut {NoStop}%
\bibitem [{\citenamefont {Urrutia}\ and\ \citenamefont
  {Castelletti}(2011{\natexlab{a}})}]{Urrutia_2011}%
  \BibitemOpen
  \bibfield  {author} {\bibinfo {author} {\bibfnamefont {I.}~\bibnamefont
  {Urrutia}}\ and\ \bibinfo {author} {\bibfnamefont {G.}~\bibnamefont
  {Castelletti}},\ }\href {\doibase 10.1063/1.3544681} {\bibfield  {journal}
  {\bibinfo  {journal} {The Journal of Chemical Physics}\ }\textbf {\bibinfo
  {volume} {134}},\ \bibinfo {eid} {064508} (\bibinfo {year}
  {2011}{\natexlab{a}})}\BibitemShut {NoStop}%
\bibitem [{\citenamefont {Reyes}(2012)}]{Reyes_2012}%
  \BibitemOpen
  \bibfield  {author} {\bibinfo {author} {\bibfnamefont {Y.}~\bibnamefont
  {Reyes}},\ }\href {\doibase 10.1016/j.fluid.2012.06.033} {\bibfield
  {journal} {\bibinfo  {journal} {Fluid Phase Equilibria}\ }\textbf {\bibinfo
  {volume} {336}},\ \bibinfo {pages} {28} (\bibinfo {year} {2012})}\BibitemShut
  {NoStop}%
\bibitem [{\citenamefont {Duda}(2009)}]{Duda_2009}%
  \BibitemOpen
  \bibfield  {author} {\bibinfo {author} {\bibfnamefont {Y.}~\bibnamefont
  {Duda}},\ }\href {\doibase 10.1063/1.3089702} {\bibfield  {journal} {\bibinfo
   {journal} {The Journal of Chemical Physics}\ }\textbf {\bibinfo {volume}
  {130}},\ \bibinfo {eid} {116101} (\bibinfo {year} {2009})}\BibitemShut
  {NoStop}%
\bibitem [{\citenamefont {Asakura}\ and\ \citenamefont
  {Oosawa}(1954)}]{Asakura_1954}%
  \BibitemOpen
  \bibfield  {author} {\bibinfo {author} {\bibfnamefont {S.}~\bibnamefont
  {Asakura}}\ and\ \bibinfo {author} {\bibfnamefont {F.}~\bibnamefont
  {Oosawa}},\ }\href {\doibase 10.1063/1.1740346} {\bibfield  {journal}
  {\bibinfo  {journal} {The Journal of Chemical Physics}\ }\textbf {\bibinfo
  {volume} {22}},\ \bibinfo {pages} {1255} (\bibinfo {year}
  {1954})}\BibitemShut {NoStop}%
\bibitem [{\citenamefont {Asakura}\ and\ \citenamefont
  {Oosawa}(1958)}]{Asakura_1958}%
  \BibitemOpen
  \bibfield  {author} {\bibinfo {author} {\bibfnamefont {S.}~\bibnamefont
  {Asakura}}\ and\ \bibinfo {author} {\bibfnamefont {F.}~\bibnamefont
  {Oosawa}},\ }\href {\doibase 10.1002/pol.1958.1203312618} {\bibfield
  {journal} {\bibinfo  {journal} {Journal of Polymer Science}\ }\textbf
  {\bibinfo {volume} {33}},\ \bibinfo {pages} {183} (\bibinfo {year}
  {1958})}\BibitemShut {NoStop}%
\bibitem [{\citenamefont {Israelachvili}(2011)}]{israelachvili_11}%
  \BibitemOpen
  \bibfield  {author} {\bibinfo {author} {\bibfnamefont {J.}~\bibnamefont
  {Israelachvili}},\ }\href@noop {} {\emph {\bibinfo {title} {Intermolecular
  and surface forces}}}\ (\bibinfo  {publisher} {Academic Press},\ \bibinfo
  {address} {Burlington, MA},\ \bibinfo {year} {2011})\BibitemShut {NoStop}%
\bibitem [{\citenamefont {Vrij}(1976)}]{Vrij_1976}%
  \BibitemOpen
  \bibfield  {author} {\bibinfo {author} {\bibfnamefont {A.}~\bibnamefont
  {Vrij}},\ }\href {\doibase 10.1351/pac197648040471} {\bibfield  {journal}
  {\bibinfo  {journal} {Pure Appl. Chem.}\ }\textbf {\bibinfo {volume} {48}},\
  \bibinfo {pages} {471} (\bibinfo {year} {1976})}\BibitemShut {NoStop}%
\bibitem [{\citenamefont {Germain}\ and\ \citenamefont
  {Amokrane}(2007)}]{Germain_2007}%
  \BibitemOpen
  \bibfield  {author} {\bibinfo {author} {\bibfnamefont {P.}~\bibnamefont
  {Germain}}\ and\ \bibinfo {author} {\bibfnamefont {S.}~\bibnamefont
  {Amokrane}},\ }\href {\doibase 10.1103/PhysRevE.76.031401} {\bibfield
  {journal} {\bibinfo  {journal} {Phys. Rev. E}\ }\textbf {\bibinfo {volume}
  {76}},\ \bibinfo {pages} {031401} (\bibinfo {year} {2007})}\BibitemShut
  {NoStop}%
\bibitem [{\citenamefont {L\'opez~de Haro}\ \emph {et~al.}(2015)\citenamefont
  {L\'opez~de Haro}, \citenamefont {Tejero}, \citenamefont {Santos},
  \citenamefont {Yuste}, \citenamefont {Fiumara},\ and\ \citenamefont
  {Saija}}]{LopezdeHaro_2015}%
  \BibitemOpen
  \bibfield  {author} {\bibinfo {author} {\bibfnamefont {M.}~\bibnamefont
  {L\'opez~de Haro}}, \bibinfo {author} {\bibfnamefont {C.~F.}\ \bibnamefont
  {Tejero}}, \bibinfo {author} {\bibfnamefont {A.}~\bibnamefont {Santos}},
  \bibinfo {author} {\bibfnamefont {S.~B.}\ \bibnamefont {Yuste}}, \bibinfo
  {author} {\bibfnamefont {G.}~\bibnamefont {Fiumara}}, \ and\ \bibinfo
  {author} {\bibfnamefont {F.}~\bibnamefont {Saija}},\ }\href {\doibase
  http://dx.doi.org/10.1063/1.4904891} {\bibfield  {journal} {\bibinfo
  {journal} {The Journal of Chemical Physics}\ }\textbf {\bibinfo {volume}
  {142}},\ \bibinfo {eid} {014902} (\bibinfo {year} {2015})}\BibitemShut
  {NoStop}%
\bibitem [{\citenamefont {Binder}, \citenamefont {Virnau},\ and\ \citenamefont
  {Statt}(2014)}]{Binder_2014}%
  \BibitemOpen
  \bibfield  {author} {\bibinfo {author} {\bibfnamefont {K.}~\bibnamefont
  {Binder}}, \bibinfo {author} {\bibfnamefont {P.}~\bibnamefont {Virnau}}, \
  and\ \bibinfo {author} {\bibfnamefont {A.}~\bibnamefont {Statt}},\ }\href
  {\doibase http://dx.doi.org/10.1063/1.4896943} {\bibfield  {journal}
  {\bibinfo  {journal} {The Journal of Chemical Physics}\ }\textbf {\bibinfo
  {volume} {141}},\ \bibinfo {eid} {140901} (\bibinfo {year}
  {2014})}\BibitemShut {NoStop}%
\bibitem [{\citenamefont {Winkler}\ \emph {et~al.}(2013)\citenamefont
  {Winkler}, \citenamefont {Statt}, \citenamefont {Virnau},\ and\ \citenamefont
  {Binder}}]{Winkler_2013}%
  \BibitemOpen
  \bibfield  {author} {\bibinfo {author} {\bibfnamefont {A.}~\bibnamefont
  {Winkler}}, \bibinfo {author} {\bibfnamefont {A.}~\bibnamefont {Statt}},
  \bibinfo {author} {\bibfnamefont {P.}~\bibnamefont {Virnau}}, \ and\ \bibinfo
  {author} {\bibfnamefont {K.}~\bibnamefont {Binder}},\ }\href {\doibase
  10.1103/PhysRevE.87.032307} {\bibfield  {journal} {\bibinfo  {journal} {Phys.
  Rev. E}\ }\textbf {\bibinfo {volume} {87}},\ \bibinfo {pages} {032307}
  (\bibinfo {year} {2013})}\BibitemShut {NoStop}%
\bibitem [{\citenamefont {Statt}\ \emph {et~al.}(2012)\citenamefont {Statt},
  \citenamefont {Winkler}, \citenamefont {Virnau},\ and\ \citenamefont
  {Binder}}]{Statt_2012}%
  \BibitemOpen
  \bibfield  {author} {\bibinfo {author} {\bibfnamefont {A.}~\bibnamefont
  {Statt}}, \bibinfo {author} {\bibfnamefont {A.}~\bibnamefont {Winkler}},
  \bibinfo {author} {\bibfnamefont {P.}~\bibnamefont {Virnau}}, \ and\ \bibinfo
  {author} {\bibfnamefont {K.}~\bibnamefont {Binder}},\ }\href
  {http://stacks.iop.org/0953-8984/24/i=46/a=464122} {\bibfield  {journal}
  {\bibinfo  {journal} {Journal of Physics: Condensed Matter}\ }\textbf
  {\bibinfo {volume} {24}},\ \bibinfo {pages} {464122} (\bibinfo {year}
  {2012})}\BibitemShut {NoStop}%
\bibitem [{\citenamefont {Noro}\ and\ \citenamefont
  {Frenkel}(2000)}]{Noro_2000}%
  \BibitemOpen
  \bibfield  {author} {\bibinfo {author} {\bibfnamefont {M.~G.}\ \bibnamefont
  {Noro}}\ and\ \bibinfo {author} {\bibfnamefont {D.}~\bibnamefont {Frenkel}},\
  }\href {\doibase http://dx.doi.org/10.1063/1.1288684} {\bibfield  {journal}
  {\bibinfo  {journal} {The Journal of Chemical Physics}\ }\textbf {\bibinfo
  {volume} {113}},\ \bibinfo {eid} {2941} (\bibinfo {year} {2000})}\BibitemShut
  {NoStop}%
\bibitem [{\citenamefont {Valadez-P\'{e}rez}\ \emph {et~al.}(2012)\citenamefont
  {Valadez-P\'{e}rez}, \citenamefont {Benavides}, \citenamefont
  {Sch\"{o}ll-Paschinger},\ and\ \citenamefont {Casta\~{n}eda
  Priego}}]{ValadezP_2012}%
  \BibitemOpen
  \bibfield  {author} {\bibinfo {author} {\bibfnamefont {N.~E.}\ \bibnamefont
  {Valadez-P\'{e}rez}}, \bibinfo {author} {\bibfnamefont {A.~L.}\ \bibnamefont
  {Benavides}}, \bibinfo {author} {\bibfnamefont {E.}~\bibnamefont
  {Sch\"{o}ll-Paschinger}}, \ and\ \bibinfo {author} {\bibfnamefont
  {R.}~\bibnamefont {Casta\~{n}eda Priego}},\ }\href {\doibase
  10.1063/1.4747193} {\bibfield  {journal} {\bibinfo  {journal} {The Journal of
  Chemical Physics}\ }\textbf {\bibinfo {volume} {137}},\ \bibinfo {eid}
  {084905} (\bibinfo {year} {2012})}\BibitemShut {NoStop}%
\bibitem [{\citenamefont {Hansen}\ and\ \citenamefont
  {McDonald}(2006)}]{Hansen2006}%
  \BibitemOpen
  \bibfield  {author} {\bibinfo {author} {\bibfnamefont {J.-P.}\ \bibnamefont
  {Hansen}}\ and\ \bibinfo {author} {\bibfnamefont {I.~R.}\ \bibnamefont
  {McDonald}},\ }\href@noop {} {\emph {\bibinfo {title} {Theory of simple
  liquids, 3rd Edition}}}\ (\bibinfo  {publisher} {Academic Press},\ \bibinfo
  {address} {Amsterdam},\ \bibinfo {year} {2006})\BibitemShut {NoStop}%
\bibitem [{\citenamefont {Blokhuis}\ and\ \citenamefont
  {Kuipers}(2007)}]{Blokhuis_2007}%
  \BibitemOpen
  \bibfield  {author} {\bibinfo {author} {\bibfnamefont {E.~M.}\ \bibnamefont
  {Blokhuis}}\ and\ \bibinfo {author} {\bibfnamefont {J.}~\bibnamefont
  {Kuipers}},\ }\href {\doibase 10.1063/1.2434161} {\bibfield  {journal}
  {\bibinfo  {journal} {The Journal of Chemical Physics}\ }\textbf {\bibinfo
  {volume} {126}},\ \bibinfo {eid} {054702} (\bibinfo {year}
  {2007})}\BibitemShut {NoStop}%
\bibitem [{\citenamefont {Paganini}(2014)}]{PaganiniTL_2014}%
  \BibitemOpen
  \bibfield  {author} {\bibinfo {author} {\bibfnamefont {I.}~\bibnamefont
  {Paganini}},\ }\href@noop {} {\enquote {\bibinfo {title} {Molecular dynamic
  simulation of colloidal particles confined in nano-cavities},}\ } (\bibinfo
  {year} {2014})\BibitemShut {NoStop}%
\bibitem [{\citenamefont {Taylor}, \citenamefont {Evans},\ and\ \citenamefont
  {Royall}(2012)}]{Taylor_2012}%
  \BibitemOpen
  \bibfield  {author} {\bibinfo {author} {\bibfnamefont {S.~L.}\ \bibnamefont
  {Taylor}}, \bibinfo {author} {\bibfnamefont {R.}~\bibnamefont {Evans}}, \
  and\ \bibinfo {author} {\bibfnamefont {C.~P.}\ \bibnamefont {Royall}},\
  }\href {http://stacks.iop.org/0953-8984/24/i=46/a=464128} {\bibfield
  {journal} {\bibinfo  {journal} {Journal of Physics: Condensed Matter}\
  }\textbf {\bibinfo {volume} {24}},\ \bibinfo {pages} {464128} (\bibinfo
  {year} {2012})}\BibitemShut {NoStop}%
\bibitem [{\citenamefont {Germain}\ and\ \citenamefont
  {Amokrane}(2010)}]{Germain_2010_b}%
  \BibitemOpen
  \bibfield  {author} {\bibinfo {author} {\bibfnamefont {P.}~\bibnamefont
  {Germain}}\ and\ \bibinfo {author} {\bibfnamefont {S.}~\bibnamefont
  {Amokrane}},\ }\href {\doibase 10.1103/PhysRevE.81.011407} {\bibfield
  {journal} {\bibinfo  {journal} {Phys. Rev. E}\ }\textbf {\bibinfo {volume}
  {81}},\ \bibinfo {pages} {011407} (\bibinfo {year} {2010})}\BibitemShut
  {NoStop}%
\bibitem [{\citenamefont {Gnan}, \citenamefont {Zaccarelli},\ and\
  \citenamefont {Sciortino}(2012)}]{Gnan_2012}%
  \BibitemOpen
  \bibfield  {author} {\bibinfo {author} {\bibfnamefont {N.}~\bibnamefont
  {Gnan}}, \bibinfo {author} {\bibfnamefont {E.}~\bibnamefont {Zaccarelli}}, \
  and\ \bibinfo {author} {\bibfnamefont {F.}~\bibnamefont {Sciortino}},\ }\href
  {\doibase 10.1063/1.4745479} {\bibfield  {journal} {\bibinfo  {journal} {The
  Journal of Chemical Physics}\ }\textbf {\bibinfo {volume} {137}},\ \bibinfo
  {eid} {084903} (\bibinfo {year} {2012})}\BibitemShut {NoStop}%
\bibitem [{\citenamefont {Hansen-Goos}, \citenamefont {Miller},\ and\
  \citenamefont {Wettlaufer}(2012)}]{HansenGoos_2012}%
  \BibitemOpen
  \bibfield  {author} {\bibinfo {author} {\bibfnamefont {H.}~\bibnamefont
  {Hansen-Goos}}, \bibinfo {author} {\bibfnamefont {M.~A.}\ \bibnamefont
  {Miller}}, \ and\ \bibinfo {author} {\bibfnamefont {J.~S.}\ \bibnamefont
  {Wettlaufer}},\ }\href {\doibase 10.1103/PhysRevLett.108.047801} {\bibfield
  {journal} {\bibinfo  {journal} {Phys. Rev. Lett.}\ }\textbf {\bibinfo
  {volume} {108}},\ \bibinfo {pages} {047801} (\bibinfo {year}
  {2012})}\BibitemShut {NoStop}%
\bibitem [{\citenamefont {Charbonneau}\ and\ \citenamefont
  {Frenkel}(2007)}]{Charbonneau_2007}%
  \BibitemOpen
  \bibfield  {author} {\bibinfo {author} {\bibfnamefont {P.}~\bibnamefont
  {Charbonneau}}\ and\ \bibinfo {author} {\bibfnamefont {D.}~\bibnamefont
  {Frenkel}},\ }\href {\doibase 10.1063/1.2737051} {\bibfield  {journal}
  {\bibinfo  {journal} {The Journal of Chemical Physics}\ }\textbf {\bibinfo
  {volume} {126}},\ \bibinfo {eid} {196101} (\bibinfo {year}
  {2007})}\BibitemShut {NoStop}%
\bibitem [{\citenamefont {Allen}\ and\ \citenamefont
  {Tildesley}(1987)}]{Allen_and_Tildesley}%
  \BibitemOpen
  \bibfield  {author} {\bibinfo {author} {\bibfnamefont {M.~P.}\ \bibnamefont
  {Allen}}\ and\ \bibinfo {author} {\bibfnamefont {D.~J.}\ \bibnamefont
  {Tildesley}},\ }\href@noop {} {\emph {\bibinfo {title} {Computer Simulations
  of Liquids}}}\ (\bibinfo  {publisher} {Clarendom Press, Oxford},\ \bibinfo
  {year} {1987})\BibitemShut {NoStop}%
\bibitem [{\citenamefont {Alder}\ and\ \citenamefont
  {Wainwright}(1959)}]{Alder_1959}%
  \BibitemOpen
  \bibfield  {author} {\bibinfo {author} {\bibfnamefont {B.~J.}\ \bibnamefont
  {Alder}}\ and\ \bibinfo {author} {\bibfnamefont {T.~E.}\ \bibnamefont
  {Wainwright}},\ }\href {\doibase 10.1063/1.1730376} {\bibfield  {journal}
  {\bibinfo  {journal} {The Journal of Chemical Physics}\ }\textbf {\bibinfo
  {volume} {31}},\ \bibinfo {eid} {459} (\bibinfo {year} {1959})}\BibitemShut
  {NoStop}%
\bibitem [{\citenamefont {Tehver}\ \emph {et~al.}(1998)\citenamefont {Tehver},
  \citenamefont {Toigo}, \citenamefont {Koplik},\ and\ \citenamefont
  {Banavar}}]{Tehver_98}%
  \BibitemOpen
  \bibfield  {author} {\bibinfo {author} {\bibfnamefont {R.}~\bibnamefont
  {Tehver}}, \bibinfo {author} {\bibfnamefont {F.}~\bibnamefont {Toigo}},
  \bibinfo {author} {\bibfnamefont {J.}~\bibnamefont {Koplik}}, \ and\ \bibinfo
  {author} {\bibfnamefont {J.~R.}\ \bibnamefont {Banavar}},\ }\href {\doibase
  10.1103/PhysRevE.57.R17} {\bibfield  {journal} {\bibinfo  {journal} {Phys.
  Rev. E}\ }\textbf {\bibinfo {volume} {57}},\ \bibinfo {pages} {17} (\bibinfo
  {year} {1998})}\BibitemShut {NoStop}%
\bibitem [{\citenamefont {Urrutia}\ and\ \citenamefont
  {Castelletti}(2011{\natexlab{b}})}]{Urrutia_11}%
  \BibitemOpen
  \bibfield  {author} {\bibinfo {author} {\bibfnamefont {I.}~\bibnamefont
  {Urrutia}}\ and\ \bibinfo {author} {\bibfnamefont {G.}~\bibnamefont
  {Castelletti}},\ }\href {\doibase 10.1063/1.3544681} {\bibfield  {journal}
  {\bibinfo  {journal} {The Journal of Chemical Physics}\ }\textbf {\bibinfo
  {volume} {134}},\ \bibinfo {eid} {064508} (\bibinfo {year}
  {2011}{\natexlab{b}})}\BibitemShut {NoStop}%
\bibitem [{Note1()}]{Note1}%
  \BibitemOpen
  \bibinfo {note} {These curves were extrapolated from surface tension data in
  Ref. \cite {LopezRendon_2006}}\BibitemShut {NoStop}%
\bibitem [{\citenamefont {Metropolis}\ \emph {et~al.}(1953)\citenamefont
  {Metropolis}, \citenamefont {Rosenbluth}, \citenamefont {Rosenbluth},
  \citenamefont {Teller},\ and\ \citenamefont {Teller}}]{Metropolis_1953}%
  \BibitemOpen
  \bibfield  {author} {\bibinfo {author} {\bibfnamefont {N.}~\bibnamefont
  {Metropolis}}, \bibinfo {author} {\bibfnamefont {A.~W.}\ \bibnamefont
  {Rosenbluth}}, \bibinfo {author} {\bibfnamefont {M.~N.}\ \bibnamefont
  {Rosenbluth}}, \bibinfo {author} {\bibfnamefont {A.~H.}\ \bibnamefont
  {Teller}}, \ and\ \bibinfo {author} {\bibfnamefont {E.}~\bibnamefont
  {Teller}},\ }\href {\doibase http://dx.doi.org/10.1063/1.1699114} {\bibfield
  {journal} {\bibinfo  {journal} {The Journal of Chemical Physics}\ }\textbf
  {\bibinfo {volume} {21}},\ \bibinfo {pages} {1087} (\bibinfo {year}
  {1953})}\BibitemShut {NoStop}%
\bibitem [{\citenamefont {Rosenbluth}\ and\ \citenamefont
  {Rosenbluth}(1954)}]{Rosenbluth_1954}%
  \BibitemOpen
  \bibfield  {author} {\bibinfo {author} {\bibfnamefont {M.~N.}\ \bibnamefont
  {Rosenbluth}}\ and\ \bibinfo {author} {\bibfnamefont {A.~W.}\ \bibnamefont
  {Rosenbluth}},\ }\href {\doibase 10.1063/1.1740207} {\bibfield  {journal}
  {\bibinfo  {journal} {The Journal of Chemical Physics}\ }\textbf {\bibinfo
  {volume} {22}},\ \bibinfo {pages} {881} (\bibinfo {year} {1954})}\BibitemShut
  {NoStop}%
\bibitem [{\citenamefont {Urrutia}\ and\ \citenamefont
  {Szybisz}(2010)}]{Urrutia_2010}%
  \BibitemOpen
  \bibfield  {author} {\bibinfo {author} {\bibfnamefont {I.}~\bibnamefont
  {Urrutia}}\ and\ \bibinfo {author} {\bibfnamefont {L.}~\bibnamefont
  {Szybisz}},\ }\href {\doibase 10.1063/1.3319560} {\bibfield  {journal}
  {\bibinfo  {journal} {Journal of Mathematical Physics}\ }\textbf {\bibinfo
  {volume} {51}},\ \bibinfo {pages} {033303} (\bibinfo {year}
  {2010})}\BibitemShut {NoStop}%
\end{thebibliography}

%

\end{document}